\documentclass[prb,superscriptaddress,twocolumn,showpacs]{revtex4}
\usepackage{graphicx}
\usepackage{times}
\usepackage{amsmath,amssymb}

\begin{document}

\title{Dynamics at and near conformal quantum critical points}

\author{S.V. Isakov}
\affiliation{Theoretische Physik, ETH Zurich, 8093 Zurich, Switzerland}

\author{P. Fendley}
\affiliation{Department of Physics, University of Virginia, Charlottesville, VA 22904-4714 USA}
\affiliation{Microsoft Research, Station Q, University of California, Santa Barbara, CA 93106} 

\author{A.W.W. Ludwig}
\affiliation{Physics Department, University of California, Santa Barbara, California 93106}

\author{S. Trebst}
\affiliation{Microsoft Research, Station Q, University of California, Santa Barbara, CA 93106} 

\author{M. Troyer}
\affiliation{Theoretische Physik, ETH Zurich, 8093 Zurich, Switzerland}

\date{\today}

\begin{abstract}

We explore the dynamical behavior at and near a special class of
two-dimensional quantum critical points. Each is a conformal quantum
critical point (CQCP), where in the scaling limit the equal-time
correlators are those of a two-dimensional conformal field theory. The
critical theories include the square-lattice quantum dimer model, the
quantum Lifshitz theory, and a deformed toric code model.  We show
that under generic perturbation the latter flows toward the ordinary
Lorentz-invariant (2+1) dimensional Ising critical point, illustrating
that CQCPs are generically unstable.  We exploit a correspondence
between the classical and quantum dynamical behavior in such systems
to perform an extensive numerical study of two lines of CQCPs in a
quantum eight-vertex model, or equivalently, two coupled deformed
toric codes.  We find that the dynamical critical exponent $z$ remains
2 along the $U(1)$-symmetric quantum Lifshitz line, while it
continuously varies along the line with only ${\mathbb Z}_2$
symmetry. This illustrates how two CQCPs can have very different
dynamical properties, despite identical equal-time ground-state
correlators.
Our results equally apply to the dynamics of the 
corresponding purely classical models.

\end{abstract}


\maketitle


\section{Introduction}

The study of critical properties of classical statistical mechanics systems
with stochastic relaxational dynamics has a venerable history.\cite{HH}  
More recently, a class of quantum systems closely related
to classical systems with stochastic dynamics has come under intense
study. Here, each basis element of the Hilbert space of the
two-dimensional quantum system corresponds to a configuration in a
two-dimensional classical system.  This in itself is not unusual; what
is special about this class is that the Hamiltonian is chosen so that
the ground state is written in terms of the Boltzmann weights of the
classical model.  One significant consequence of this relation between
quantum and classical models is that correlators in the ground state
of such a Hamiltonian are the same as those of the classical
model. This means that the phase diagrams of the quantum and classical
models are closely related.

The quantum dimer model \cite{RK} is an example of this class of
quantum systems. On the square lattice, it is a canonical example of
the conformal quantum critical points to be discussed in this paper,
while on the triangular lattice is a canonical example of topological
order.\cite{MS} The Hilbert space is spanned by close-packed
hard-core dimer configurations. The quantum ground state is the
equal-amplitude sum over all such dimer configurations. A Hamiltonian
having this ground state can easily be constructed by using a
technique first developed in this context by Rokhsar and Kivelson \cite{RK}
(RK). It typically is a sum of local projection operators, each
annihilating the ground state, and will be reviewed in Sec.
\ref{qdm}.

In particular, this provides a way to discover and analyze new quantum
critical points in two spatial dimensions. Whenever a ground state in two
dimensions has a correlator of local operators algebraically decaying
with distance, all local Hamiltonians must be gapless.\cite{Hastings} 
Thus if the ground state of an RK Hamiltonian is
described by a critical classical theory, the quantum theory must be
critical as well. Since well-understood quantum critical points in two
dimensions are few and far between, such RK Hamiltonians are quite
interesting.\cite{FootnoteFrustration}

Indeed some of the best-understood quantum critical points in two dimensions
were found by analyzing RK Hamiltonians. A great deal is known about
critical points in rotationally invariant two-dimensional classical
models, because these critical points are not only scale invariant,
but conformally invariant as well.  Two-dimensional conformal symmetry
has an infinite number of generators, and so is very powerful. It can
be used not only to identify many classes of critical points, but to
explicitly compute correlation functions in the scaling limit, even
when the systems are strongly interacting. Because the ground state
correlators are those of the classical theory, these results can then
be carried over to the quantum case. In fact, the ground state itself 
 becomes conformally invariant in the scaling limit,
as detailed in Ref.\ \onlinecite{AFF}.
For this reason, these theories were
dubbed {\em conformal quantum critical points}.

The static behavior of conformal quantum critical points is well
understood because of this connection to conformal field theory.
Their {\em dynamical behavior}, however, is another story. Only in an
exactly solvable case \cite{Henley97,AFF,Senthil05}, dubbed
the quantum Lifshitz theory, has the quantum
dynamics has  been studied in depth. Here the ground-state correlators
can be written in terms of classical correlators of a free massless
scalar field. Moreover, the quantum Hamiltonian remains quadratic (although
not Lorentz-invariant), and so everything in principle can be computed.\cite{Senthil05}

One purpose of this paper is to study the dynamics of more complex
conformal quantum critical points. A key example, which we will study
in detail, are the two lines of conformal quantum critical points in the
quantum eight-vertex model of Ref.\
\onlinecite{AFF}. 
The first of these two lines exhibits an additional $U(1)$ symmetry, 
which reduces the quantum eight-vertex model to a quantum six-vertex 
model whose scaling limit  corresponds precisely to the quantum Lifshitz 
model mentioned above. So, while the quantum six-vertex model is not exactly
solvable, the field theory describing its scaling limit is. 
There is, however, an RK Hamiltonian whose ground state is that
of the classical six-vertex model. The second critical line corresponds to the full
quantum eight-vertex model. This line is not exactly solvable
on the lattice or in the scaling limit, although again the ground
state of an RK Hamiltonian can be found explicitly.  The equal-time
correlators on the two critical lines can be mapped onto each other,
because the two corresponding classical theories are dual to each
other.\cite{Baxbook}

In general, different dynamical symmetry classes
are possible for a given static universality class, depending, e.g.,
on whether the dynamics possesses conservation laws or not.\cite{HH}
We find here that despite the identical equal-time correlators, the
dynamical behavior of the two critical lines
is indeed quite different. In the quantum Lifshitz theory, the dynamical
critical exponent remains $z=2$ all along the line, while along the
other line, this exponent is found to vary. 
In the language of Ref.\ \onlinecite{HH}, the latter
possesses {\it Model A} dynamics, 
i.e.\  the dynamics has no conservation law, and only
Ising ${\mathbb Z}_2$ symmetry. In the former, however, 
the dynamics respects a $U(1)$ symmetry, and so 
has a conserved quantity; this corresponds to {\it Model B}.
Both theories possess an exactly marginal operator, and the
exponents of the equal-time correlators vary continuously along the critical
lines. However, we note that the presence of an exactly marginal operator implies
the existence of a $U(1)$ symmetry only in two classical dimensions,
not in a full two-dimensional quantum theory.

To do this analysis, we exploit the fact that the connection between
the classical and quantum models goes deeper than the equal-time
correlators. As described in refs.\
\onlinecite{Henley97,Henley04,Castelnovo05}, much can be learned about
the quantum dynamics by studying the classical ones. This means in
particular that in some cases the dynamical critical exponent $z$ in
the quantum model can be found by doing classical Monte Carlo
simulations. Since one point along the $z\ne 2$ line amounts to doing
classical Ising dynamics, we make contact with decades of numerical
studies \cite{LandauBinder} here.

Even though an RK Hamiltonian is fine-tuned, when it describes a phase
with topological order, this physics persists under (at least small)
deformations. This has been demonstrated numerically in several
examples,\cite{MS,Trebst07,Vidal09, Tupitsyn10,Vidal10} and recently been derived rigorously.
\cite{Klich,Bravyi10} The robustness of topological order near RK
points is not surprising, given that the phase is gapped. The argument
of course does not apply at conformal quantum critical points, and
there is no reason to expect that these will remain critical under
generic perturbations. Even though the square-lattice quantum dimer
model is critical with seemingly no fine tuning, this is a consequence
of the highly constrained behavior of dimer models. In a more general
setting, it has been shown that this quantum critical point typically
has several relevant perturbations, as well as a host of dangerously
irrelevant operators.\cite{Fradkin03,Balents03}

Thus an interesting question is if a conformal quantum critical
point is isolated, or if it is part of a phase boundary. In
renormalization group language, the question is whether a relevant
perturbation causes a flow to another quantum critical point, or into
a gapped phase. We will present substantial evidence that the Ising
conformal quantum critical point is continuously connected to the
usual 2+1d Ising critical point.

The outline of the remainder of the paper is as follows.  In Sec.
\ref{cqcpRK} we review some well known models with RK Hamiltonians,
the quantum Lifshitz theory and the quantum dimer model. We also
display a general connection between classical dynamics and quantum
dynamics in theories with RK Hamiltonians.  In Sec.
\ref{SectionQuantumClassicalDynamicsIsing}, we discuss the
dynamics of the Ising conformal quantum critical point in two spatial
dimensions, i.e.\ the quantum critical point whose ground state is
written in terms of the Boltzmann weights of the critical classical 2d
Ising model. RK Hamiltonians are strongly fine-tuned, and so we show
in Sec. \ref{Flow2Dto3DIsing} that generic perturbations of the
lattice model generate a crossover from the $d=2$ dimensional 
dynamics with dynamical critical exponent $z \approx 2.167$ (and 2D
static correlation length exponent $\nu =1$), to the critical dynamics
of the (2+1)-dimensional classical Ising universality class with $z=1$
(and $\nu \approx 0.632$).  In Sec. \ref{sec:cqcp} we discuss the
case of the dynamics of {\it two coupled} copies of the deformed toric
code, or in the equivalent classical case, two coupled copies of the
critical Ising model.
There are two critical lines,
whose equal-time correlators can be mapped into each other's by
two-dimensional classical duality. We show that these lines correspond
to different dynamic universality classes.  In the case without $U(1)$
symmetry, we will present numerical results that indicate that along
with the static critical exponents, the dynamical critical exponent
$z$ also varies with the couplings.


\section{Conformal quantum critical points and RK Hamiltonians}
\label{cqcpRK}


The great progress made in understanding conformal quantum points came
from considering specific ground state wavefunctions. Once
a particular ground state is specified, it is usually (but not always)
straightforward to find some RK Hamiltonian with this ground state.

All the models we study have Hilbert spaces whose basis elements are
labeled by configurations in some classical two-dimensional model.
Let ${\cal C}$ label a classical configuration, and $w({\cal C})$ be
its Boltzmann weight. Then let $|{\cal C}\rangle$ be the corresponding
basis element of the quantum Hilbert space. Here we take the simplest
choice for the inner product, the orthonormal one. In a lattice model
with discrete degrees of freedom,
\begin{equation}
\langle {\cal C}'|{\cal C}\rangle = \delta_{C C'}\ .
\label{OrthoNormal}
\end{equation}
With continuous degrees of freedom as in a field theory, the Kronecker
delta is replaced by a Dirac delta function. 

To find a conformal quantum critical point (CQCP) by this method, the
classical model must be critical and isotropic, and the Boltzmann
weights $w({\cal C})$ must be real and non-negative for all ${\cal
  C}$. Then the (unnormalized) ground state wave function
\begin{equation}
|\Psi\rangle = \sum_{\cal C} \sqrt{w({\cal C})} |{\cal C}\rangle
\label{gs}
\end{equation}
is that of a CQCP. The expectation value of any diagonal operator
${\cal D}$ in the ground state is identical to that found in the
classical theory:
$$ \frac{\langle\Psi|{\cal D}|\Psi\rangle}{\langle\Psi|\Psi\rangle} = 
\frac{\sum_{\cal C} w({\cal C}) {\cal D}({\cal C})}
{\sum_{\cal C} w({\cal C})}\ ,$$
where ${\cal D}({\cal C })\equiv \langle{\cal C}|{\cal D}|{\cal
  C}\rangle$ is by definition the value of ${\cal D}$ in
configuration ${\cal C}$.

At first glance, it appears that for any local dynamics possible in a 
critical classical system, there exists a corresponding conformal
quantum critical point. This is not true, because in a quantum theory
one of course weights by the absolute value of the amplitude
squared. Some classical Euclidean field theories like the
Wess-Zumino-Witten models are only critical when the action includes
an imaginary term. Thus even if one has an RK Hamiltonian whose
ground state includes such a term, this term disappears from
$|\Psi|^2$, and so the correlators will not decay algebraically.\cite{AFF}

\subsection{The quantum Lifshitz theory}
\label{Sec:Lifshitz}

An important example of a CQCP is given by the quantum Lifshitz
theory.  This is an exactly solvable field theory that describes,
among other things, the scaling limit of the square-lattice quantum
dimer model. Here the degrees of freedom are given by a scalar field
$\varphi(x)$, which is a smooth map of all points $x$ on some
two-dimensional manifold 
to a circle. 
The Boltzmann weight of the classical model for a given field
configuration is simply $w=e^{-S_{2d}}$, where $S_{2d}$ is the
standard action for a free massless scalar field:
\begin{equation}
S_{2d}(\varphi) = \kappa \int d^2 x (\nabla\varphi)^2 \ ,
\label{S2d}
\end{equation}
with $\kappa$ an arbitrary parameter. As discussed in Ref.\
\onlinecite{AFF}, this weight defines a ground-state wave functional,
such that all diagonal correlators in the quantum model are those of a
classical massless scalar field in two dimensions.

The action $S_{2d}$ is invariant under conformal transformations of
$x$, so the Boltzmann weights for the quantum Lifshitz ground state are
invariant as well. It also possesses an additional symmetry, under
shifts (constant in space) of $\varphi$. This $U(1)$ symmetry is also
a symmetry of the quantum Lifshitz Hamiltonian
\begin{equation}
H_{QL}=\frac{1}{2} \int d^2x \left[ \Pi^2 + \kappa^2
  (\nabla^2\varphi)^2\right]\ ,
\label{qlham}
\end{equation}
where the canonical conjugate field $\Pi(x)$ obeys
$[\Pi(x),\varphi(x')] = i \delta^{(2)}(x-x')$. 
For the examples of interest here, the field $\varphi $ is periodic, i.e.\ $\varphi = \varphi+2\pi$. 
The resulting vortex operators, however, are irrelevant at a CQCP.\cite{AFF}

This Hamiltonian is not a sum of projection operators like the RK lattice Hamiltonian
is. However, it is very similar: it can be written \cite{AFF} as the
integral over local terms of the
form $Q^\dagger(x)Q(x)$. Such a Hamiltonian necessarily has
  non-negative eigenvalues, and so any state annihilated by $Q(x)$ for
  all $x$ is a zero-energy ground state. The ground state (\ref{gs})
  with $w=e^{-S_{2d}}$ is indeed annihilated by all $Q(x)$.

Since the quantum Lifshitz Hamiltonian is quadratic in the fields, it is
exactly solvable, and so much more than just the ground-state
correlators can be computed.\cite{Senthil05} In particular, by
writing down the three-dimensional Euclidean action for this theory
\begin{equation}
S_{3d}= \frac{1}{2} \int d^2x dt \left[
  \left(\frac{\partial\varphi}{\partial t}\right)^2 + \kappa^2
  (\nabla^2\varphi)^2\right]\ ,
\label{S3d}
\end{equation}
one sees immediately that the dispersion relation is $E\propto k^2$,
and so the dynamical critical exponent is $z=2$. As we will discuss
below, we believe that a generic $U(1)$-invariant CQCP has $z=2$.

\subsection{Quantum dimers on the square lattice}
\label{qdm}

The square-lattice quantum dimer model is an example of a lattice
model with a conformal quantum critical point.\cite{RK} Each basis
element ${\cal C}$ corresponds to a configuration of dimers stretching
between neighboring sites on the square lattice. The weight $w({\cal
C})=1$ for all configurations that have exactly one dimer touching
each site, and is zero for all configurations violating the
constraint. It has long been known that correlators in the classical
dimer model on the square lattice are algebraically decaying.\cite{FS} 
Thus by Hastings' theorem,\cite{Hastings} the square-lattice quantum dimer
model is critical.

The RK Hamiltonian here was described in the original paper by
Rokhsar and Kivelson.\cite{RK} It is a sum of local projection
operators with the ground state annihilated by each projector
individually. Each term acts non-trivially on a single plaquette,
projecting each of the two configurations with two dimers on this
plaquette onto their difference. It annihilates all other
configurations on that plaquette. Each such projector indeed
annihilates the equal-amplitude sum over all configurations.

An elegant argument \cite{Henley97} indicates that the scaling limit 
of the square-lattice quantum dimer model is $\kappa^{-1}=2\pi$ of the
quantum Lifshitz model: First, one identifies the
classical model as having the free scalar field action $S_{2d}$ in
(\ref{S2d}) with $\kappa^{-1}=2\pi$. This follows from a Coulomb-gas
mapping, done here by rewriting the dimer degrees of freedom in terms
of a classical ``height'', an integer-value degree of freedom on every
site of the lattice. It is then natural to assume that the height
becomes the continuous field $\varphi$ in the scaling limit. The value
$\kappa$ is then identified by comparing the scaling dimensions
determined from the exact computation to those in the classical 2d
scalar field theory.  The ground state in the scaling limit therefore must be
of the form (\ref{gs}), because all its diagonal correlators are the
same as those of $S_{2d}$. The Hamiltonian (\ref{qlham}) by
construction annihilates this. Moreover, note that the quantum dimer
model has a $U(1)$ symmetry. The Hamiltonian preserves both the
overall number of dimers and their ``winding number'' of
dimers.  (The winding number is defined by $\sum_j n_j(-1)^{j}$, where
$j$ labels the links around a cycle and $n_j=1$ for occupied links and
$n_j=0$ for unoccupied.)  This requires that only derivatives of
$\varphi$ can appear in the Hamiltonian, which indeed is a property of
(\ref{qlham}).

\begin{figure}
\includegraphics[width=\columnwidth]{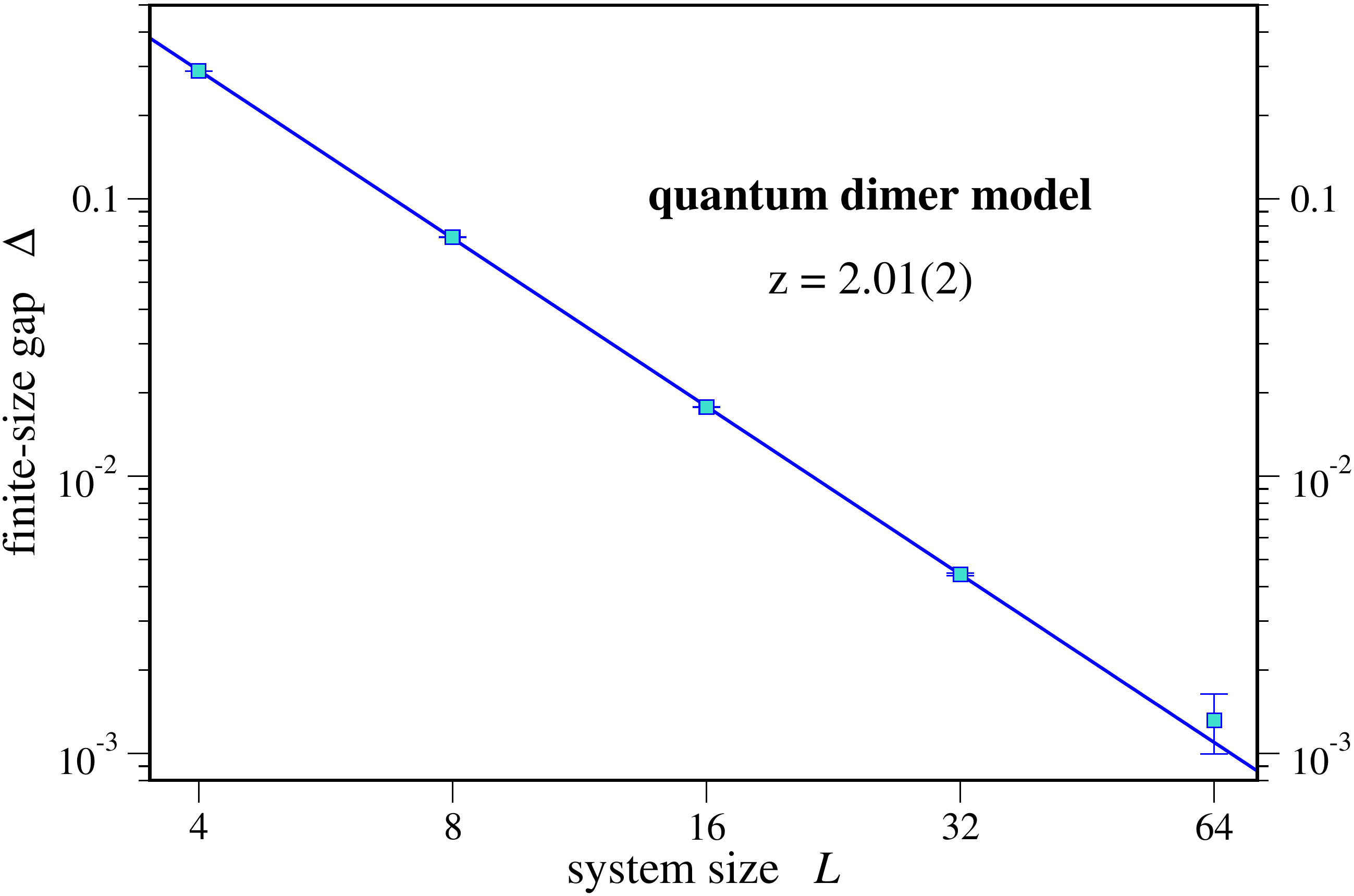}
\caption{The finite-size gap of the quantum dimer model as a function of the system
size. The gap scales as $\Delta \propto L^{-2.01(2)}$ indicating that
$z=2.01(2)$.}
\label{fig:dimer:gap}
\end{figure}
Finding the dynamical scaling exponent $z$ provides another check that
the scaling limit of the square-lattice quantum dimer model is given
by the quantum Lifshitz theory.  The quantum dimer model is not
solvable, even though its ground state is known exactly. We thus need
to find $z$ numerically; we do this by studying the finite-size
scaling of the gap. Our method is described in Appendix \ref{Sec:Numerics}.
We find
that $z=2.01(2)$, as shown in Fig.~\ref{fig:dimer:gap}. This is in excellent
agreement with the exact value $z=2$ for the quantum Lifshitz theory,
as seen from (\ref{S3d}), and previous numerical results.\cite{Laeuchli08}
As such it also provides a nice consistency check on our numerical method.

\subsection{RK Hamiltonians from classical stochastic dynamics}
\label{sec:stochastic}

For classical systems with (positive) Boltzmann weights 
$w({\cal C})$, 
every relaxational stochastic dynamics is known\cite{Henley04}
to give rise to a ``RK Hamiltonian'': 
time-dependent probabilities $p_{\cal C}(\tau) \geq 0 $,
satisfying a ``master equation''. 
The master equation is written in terms of a ``transition matrix''
$W_{{\cal C}, {\cal C}'}$, where 
for different configurations
${\cal C} \not = $ ${\cal C}'$,
the matrix element
$W_{{\cal C}, {\cal C}'} \geq 0$ denotes the transition
rate from
${\cal C}'$
to
${\cal C}$.
The diagonal element defined by
$W_{{\cal C}, {\cal C}} :=$
$- \sum_{{\cal C}' \not = {\cal C}}
W_{{\cal C}', {\cal C}}$ denotes the transition
rate out of state ${\cal C}$. The master equation is then
\begin{equation}
\label{MasterEq}
\frac{d}{ d\tau} p_{\cal C}(\tau)
=
\sum_{{\cal C}'} \ W_{{\cal C}, {\cal C}'} \ p_{{\cal C}'}(\tau)\ .
\end{equation}
The probabilities
relax at long times to the classical equilibrium distribution
\begin{equation}
p_{\cal C}(\tau) \ \to  w({\cal C}) 
\label{RelaxToEquil}
\end{equation}
when detailed balance
\begin{equation}
 w({\cal C}) \ W_{{\cal C}', {\cal C}} 
\ 
=
\
W_{{\cal C}, {\cal C}'} \ w({\cal C}')
\label{DetailedBalance}
\end{equation}
is satisfied.

A slight re-writing of the master equation (\ref{MasterEq})
gives a generalization of the RK Hamiltonian.
The rescaled transition matrix
\begin{equation}
{\widetilde W}_{{\cal C}, {\cal C}'} \ 
\equiv
\frac{1} {\sqrt{w({\cal C})} }
\
W_{{\cal C}, {\cal C}'} \
\sqrt{w({\cal C}')}\ ,
\end{equation}
is (real) symmetric due to detailed balance Eq.\
(\ref{DetailedBalance}).  Letting $ {\tilde p}_{\cal C}(\tau) \equiv
p_{\cal C}(\tau)/ \sqrt{w({\cal C})} $, the rewritten master equation
is
\begin{equation}
\label{MasterEqRewrite}
\frac{d}{ d\tau} 
{\tilde p}_{\cal C}(\tau) 
=
\sum_{{\cal C}'} \ {\widetilde W}_{{\cal C}, {\cal C}'} \ 
{\tilde p}_{{\cal C}'}(\tau)  \\
\end{equation}
This is a Schr\"odinger equation in imaginary time $\tau$ 
with Hamiltonian
${H}_{{\cal C}, {\cal C}'}=  
(-1) {\widetilde W}_{{\cal C}, {\cal C}'}$  
for the time-dependent quantum state
\begin{equation}
|\Psi(t)\rangle = \sum_{\cal C} 
\
{\tilde p}_{\cal C}(t)
\  |{\cal C} \rangle,
\label{TimeDependentState}
\end{equation}
Note that
$ {\widetilde W}_{{\cal C}, {\cal C}'}$ 
and
$ {W}_{{\cal C}, {\cal C}'}$ 
are related by a similarity transformation and so
have the same spectrum.
The Hamiltonian thus has the same spectrum as 
that of the
relaxational dynamics, Eq.\ (\ref{MasterEq}).
The state (\ref{TimeDependentState}) 
relaxes, due to Eq.\ (\ref{RelaxToEquil}),  
at long imaginary times $\tau$ to the ground state Eq.\ (\ref{gs}).
Thus this construction indeed 
provides a generalization of the RK-type Hamiltonian and ground state.

In the present formulation, ${\cal C}$ denotes configurations of
classical variables on a lattice. When the system with classical
Boltzmann weights $w({\cal C})$ possesses an equilibrium critical
point, a universal continuum field theory description emerges upon
coarse graining for both the static as well as for the dynamic
correlations.  The resulting dynamic universality classes are
classified within the well known framework of Hohenberg and Halperin.
\cite{HH} For example, the Ising conformal quantum critical point
discussed below in section \ref{SectionQuantumClassicalDynamicsIsing}
is described by the { Model A} dynamical universality class of
Hohenberg and Halperin\cite{HH} (describing dynamics lacking any
conservation law). The quantum Lifshitz universality class, however,
possesses a $U(1)$ symmetry, so that this belongs to the {Model B}
universality class.

The classical dynamics are conventionally described by a time-dependent
Landau-Ginzburg equation with stochastic Langevin-type noise, or
equivalently, as a Fokker-Planck equation for the time-evolution of
the probability distribution for the coarse-grained classical degrees
of freedom.  It is well known\cite{ZinnJ} that if one performs the
corresponding similarity transformation to make the linear operator
appearing in the Fokker-Planck equation Hermitian, the Fokker-Planck
equation turns into the coarse-grained analog of the Schr\"odinger
equation, Eq.\ (\ref{MasterEqRewrite}), in imaginary time.
Specifically, for a classical statistical mechanics system in $d$
spatial dimensions described, e.g., by a static (real) Landau-Ginzburg
action $S{\{\phi_a(x)\}}$ for coarse-grained classical degrees of
freedom $\{\phi_a(x)\}$,
the resulting quantum Hamiltonian is again of the form
\begin{equation}
H = \frac{1}{ 2}\int d^d x \ Q_a^\dagger(x) Q_a(x)
\label{HamiltonianFokkerPlanck}
\end{equation}
generalizing Eq.\  (\ref{qlham}),
where
\begin{eqnarray}
\nonumber
Q_a(x) &=&
\frac{1}{ \sqrt{2}}
\left (
\frac{\delta}{\delta \phi_a(x)} +\frac{1}{2} \frac{\delta S}{\delta \phi_a(x)}
\right )\\
\label{DEFQQdagger}
Q^\dagger_a(x) &=&
\frac{1}{ \sqrt{2}}
\left (
-\frac{\delta}{\delta \phi_a(x)} + \frac{1}{2}  
\frac{\delta S}{\delta \phi_a(x)}
\right )
\end{eqnarray}
This 
implies very generally 
that the ground state wave function
 $\Psi(\{ \phi_a (x) \} ) =$
$ (1/\sqrt{Z}) \exp \{ -\frac{1}{2} S{ \{\phi_a(x)\}} \}$
is given by the classical Boltzmann weight
($Z$ denotes the classical partition function).
Thus the (in general non-Hermitian)
linear operator appearing in the Fokker-Planck equation 
becomes, after similarity transform, the negative
of the (Hermitian) quantum Hamiltonian, $(-1) H$. All its
eigenfunctions are positive.

The Hamiltonians of the form of Eq.\  (\ref{HamiltonianFokkerPlanck})
are clearly very special.  One may ask
if there exists a  symmetry principle which constrains Hamiltonians
to be fine-tuned to this very particular form.  Indeed, there exists
such a symmetry.  It is again well 
known\cite{ZinnJ} that classical relaxational stochastic dynamics
can be viewed as being invariant under  certain supersymmetry
transformation.\cite{DijkgraafOrlandoReffert}
The property that equal-time correlation functions
converge at large times to static equilibrium correlation function
can then be seen as a direct consequence of the underlying supersymmetry.
The perturbation of the Ising conformal quantum critical point
discussed in Sec. \ref{Flow2Dto3DIsing} can thus be
viewed as breaking the supersymmetry.

This procedure makes obvious the relation between classical dynamics
and RK Hamiltonians. In fact, the generic way of constructing
RK Hamiltonians in field theory discussed in Ref.\ \onlinecite{AFF} is
formally identical to that in Eqs.\
(\ref{HamiltonianFokkerPlanck}) and (\ref{DEFQQdagger}). For example, the
quantum Lifshitz Hamiltonian (\ref{qlham}) arises with one field
$\varphi$ and setting $S=S_{2d}$ from Eq.\ (\ref{S2d}).\cite{Henley97,AFF}

\section{Quantum and classical dynamics in the Ising model}
\label{SectionQuantumClassicalDynamicsIsing}

The Hilbert spaces of the lattice models we consider for the remainder
of the paper are simply those of the Ising model on the square
lattice. Namely, at every site we have a spin (i.e.\ a two-state
quantum system), and all the Hamiltonians we consider are invariant
under flipping all the spins. We write the Hamiltonians in terms of
the Pauli matrices $\tau^a_\alpha$ acting on the spin at site $\alpha$.

\begin{figure}[t]
\includegraphics[width=.8\columnwidth]{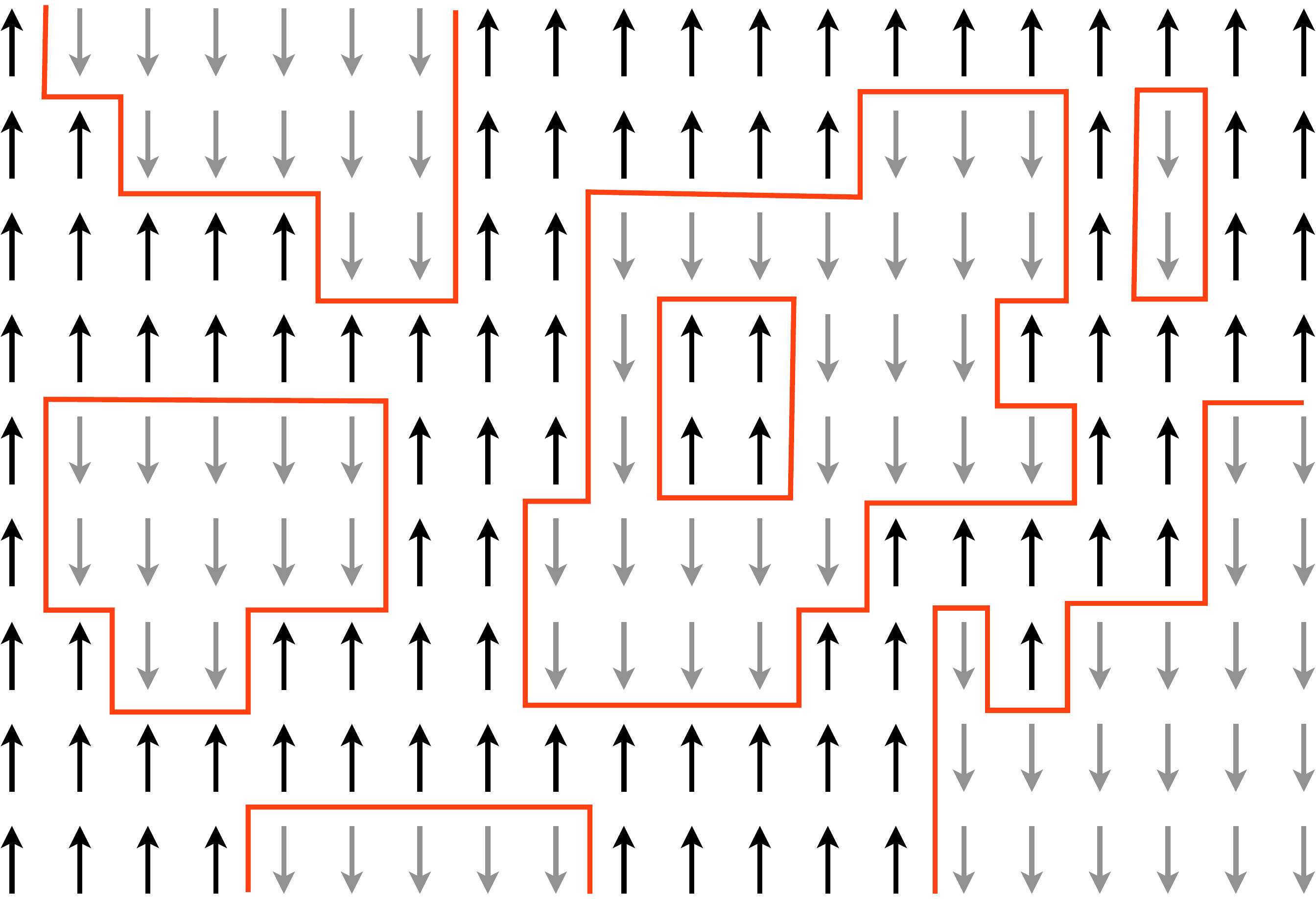}
\caption{Sketch of a typical domain wall configuration separating regions of opposite spins.}
\label{fig:loops}
\end{figure}

For both pedagogical and physical reasons, it is often convenient to
describe the Hilbert space in terms of {\em domain walls} instead of
spins. There is a two-to-one mapping from spin configurations to
domain walls obtained simply by drawing a lines on the links of the
dual lattice separating regions of different spins. A typical
configuration is illustrated in Fig.~\ref{fig:loops}.
Each domain wall is a
two-state system, so we define the Pauli matrices $\sigma^a_k$ acting
on the link $k$ analogously to the spins. However, the domain walls
cannot end or branch, so there must be an even number of domain walls
at each site. In an equation, we constrain
\begin{equation}
\prod_{k \hbox{ touching }\widehat{\alpha}} \sigma^z_k=1\ ,
\label{constraint}
\end{equation}
 for all sites $\widehat\alpha$ on the dual lattice.
For the obvious reason, these domain walls are often referred to as
``loops''.

\subsection{The toric-code ground state}

Let us first consider an important non-critical case, closely related
to the mother of all lattice models for topological quantum
computation, the toric code.\cite{Kitaev97} Here the ground state
(\ref{gs}) is a sum over all Ising configurations with the same
amplitude $\sqrt{w({\cal C})}=1$. Thus the associated classical model
is simply the Ising model at infinite temperature. Equivalently, the
ground state can be viewed as a sum over all loop configurations with
equal amplitude. Correlators of local objects in the ground state
(e.g.\ whether a link is occupied by a loop or not) are obviously
finite-range.

The gapped RK Hamiltonian having this ground state is simply
the sum over all possible single-spin flips:
\begin{equation}
H_{T=\infty}= \sum_\alpha (1-\tau^x_\alpha)\ .
\label{flip}
\end{equation}
In terms of domain
walls/loops, this acts on the four links on the dual lattice
surrounding each site $j$, removing a wall when there is one and
adding a wall when there isn't:
\begin{equation}
\tau^x_\alpha = \prod_{p\in \square_\alpha} \sigma^x_{p}
\label{tausigma}
\end{equation}
where the product is over the four links on the plaquette on the dual lattice
surrounding the site $\alpha$. 

This equal-amplitude sum over all loops/domain walls is the same
ground state as in the toric code. The difference between the models
is that in the toric code, the constraint (\ref{constraint}) is not
imposed. Rather, a diagonal term is added to the Hamiltonian to give
an energy for every sites with an odd number of domain walls touching
it. The energy penalty then precludes any configuration with an odd
number of domain walls at a site from appearing in the ground state.

Since each term in the Hamiltonian (\ref{flip}) commutes with each of
the others, this model is trivially solvable, and the sum over all
loops is indeed the ground state. On the disk, this ground state is
unique. On the torus, there are four ground
states, corresponding to the choices of having zero or one loops go
around each cycle. It is easy to see that the model is gapped; excited
states correspond to having plaquettes with the eigenvalue $-1$ of
$\tau^x$. This ground state has topological order, and so has been the
subject of intense study in recent years.

\subsection{The Ising CQCP}

Thinking about the toric code in terms of Ising spins makes it clear
how to find a conformal quantum critical point separating the phase
with topological order from an ordered phase. We simply need to change
the weight in the ground state to correspond to that in
the classical two-dimensional Ising model at non-infinite
temperature. In loop language, this corresponds to adding a weight per
unit length $2\eta$, so that
$$w({\cal L})= \eta^{2n_{\cal L}}$$
where $n_{\cal L}$ is the number of links covered by loop. 
In the
quantum theory, we can write
$$ \left(\sum_{k=1}^N \sigma^z_k \right) |{\cal L}\rangle 
= (2n_{\cal L}-N) |{\cal L}\rangle$$
for $N$ links.
The ground state is then simply
\begin{equation}
 |\Psi \rangle = \sum_\mathcal{L} \eta^{n(\mathcal{L})}
  |\mathcal{L}\rangle\ .
\label{icqcp}
\end{equation}

An RK Hamiltonian with the ground state (\ref{icqcp}) for general
$\eta$ can be found by including a potential that weights links. We
will show below that the Hamiltonian found in Ref.\ \onlinecite{AFF}
gives two copies; a simpler-to-write form is \cite{CC}
\begin{equation}
H_{Ising} = \sum_\alpha\left(\prod_{p\in \square_\alpha} \eta^{\sigma^z_p} -
\prod_{p\in \square_\alpha} \sigma^x_{p}\right)\  .
\label{Hising}
\end{equation}
This Hamiltonian can be written as a sum over projectors, each
projector breaking into two-by-two blocks. Note that there is no
manifest $U(1)$ symmetry like there is in the quantum dimer model.

The 2d Ising critical point is at $\eta=\eta_c$, where
\begin{equation}
\eta_c = (1+\sqrt{2})^{-1/2}
\label{etac}
\end{equation}
Any local Hamiltonian such as (\ref{Hising}) having (\ref{icqcp}) with
$\eta=\eta_c$ as a ground state must be quantum critical by Hastings'
theorem.  We call the Hamiltonian (\ref{Hising}) with $\eta=\eta_c$
the {\em Ising CQCP}.

This quantum Hamiltonian imposes a dynamics on the spins {\em
identical} to the usual classical dynamics with local updates.  Thus
the dynamical exponent $z$ is the same in two cases. There have been
numerical determinations of $z$ for decades; the current value from
Ref.~\onlinecite{Nightingale} is $z=2.1667(5)$.  This is slightly larger than
the $z=2$ value for the quantum dimer model and the quantum Lifshitz
theory; below in Sec.  \ref{sec:cqcp} we explain how to adapt an
argument of Ref.\ \onlinecite{AFF} to suggest that
$z\ge 2$ for CQCPs.  For the Ising case here, a rigorous inequality
requires that $z\ge 7/4$.\cite{Halperin73}

\begin{figure}
\includegraphics[width=\columnwidth]{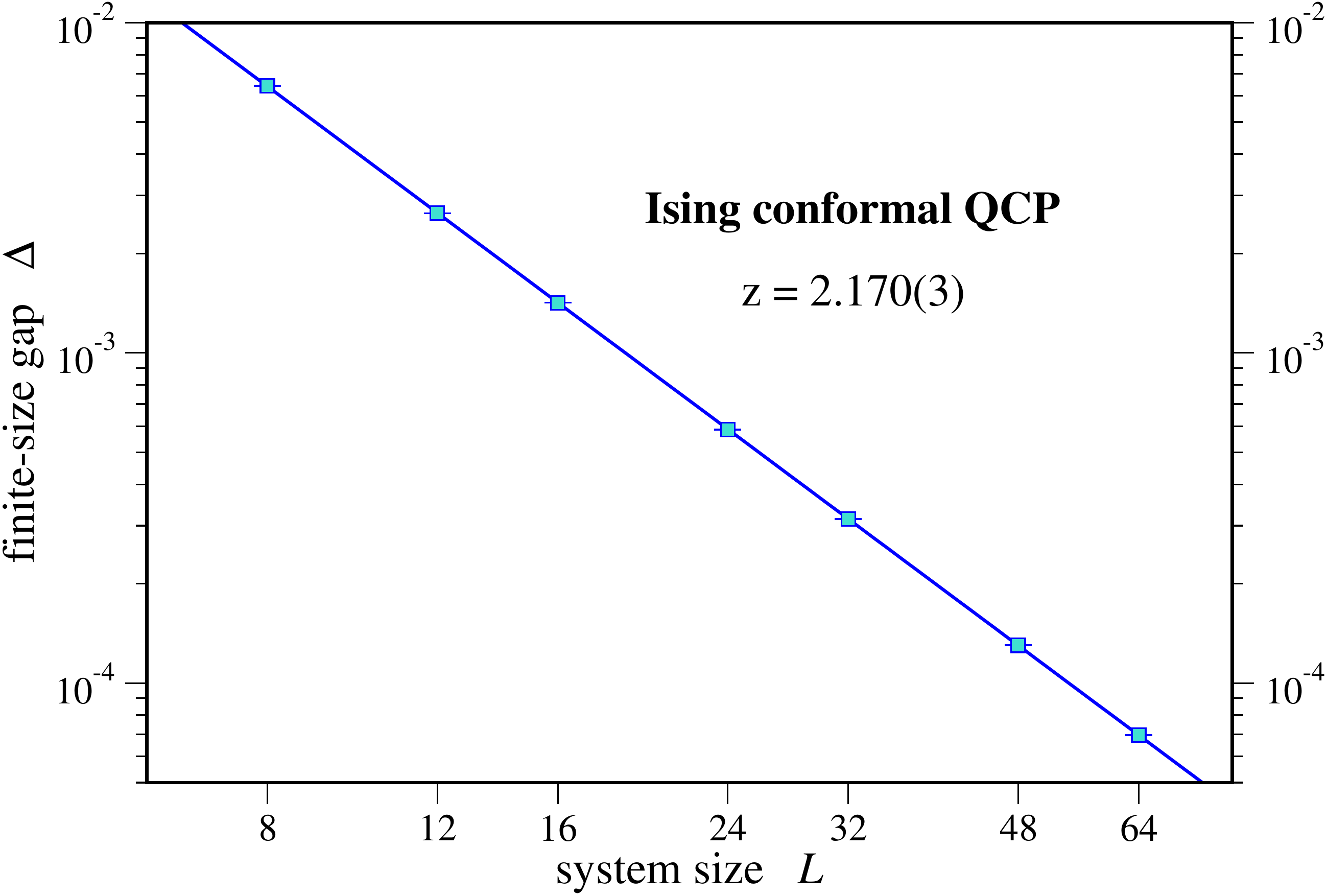}
\caption{The finite-size gap of Hamiltonian~\eqref{Hising} at the Ising CQCP $\eta_c$
as a function of the system size. The gap scales as
$\Delta \propto L^{-2.170(3)}$ indicating that $z=2.170(3)$.}
\label{fig:ising:gap}
\end{figure}

We have rechecked this number using our numerical methods described in
Appendix \ref{Sec:Numerics}, and find the same result, see
Fig.~\ref{fig:ising:gap}. We note that one needs to simulate the
classical Ising model with the Monte Carlo transition matrix that is
proportional to the quantum Hamiltonian.  In this case, the transition
matrix has the following form
$$
  W_{{\cal C},{\cal C}'} =
    \eta^4 \sqrt{\frac{w({\cal C}')}{w({\cal C})}},
$$
where $w({\cal C})$ and $w({\cal C'})$ are the weights of the configurations
before and after update (single spin flip). This transition matrix leads to
the same dynamical exponent as the Metropolis transition matrix.


\section{Flow from 2D Ising to 3D Ising}
\label{Flow2Dto3DIsing}

Another interesting way to deform the toric code is to include the
more conventional Ising nearest-neighbor interaction.  In this
section, we study the phase diagram of the toric code deformed by this
interaction and the special RK-type interaction in (\ref{Hising}). We
will show that including this term causes a flow from the Ising CQCP with
2d Ising critical exponents to the usual $z=1$ Ising
critical transition in $2+1$ dimensions.

Adding the Ising nearest-neighbor interaction deforms
Eq.~(\ref{Hising}) to
\begin{equation}
 H = -\frac{1}{\eta^4} \sum_{\alpha} \left[
\tau_{\alpha}^x
     +\frac{1}{\eta^4} 
	    \prod_{\beta}\eta^{\tau_{\alpha}^z \tau_{\beta}^z}
	 -\frac{h\eta^4}{2} \sum_{\beta} \tau_{\alpha}^z\tau_{\beta}^z\right],
\label{eq:h2}
\end{equation}
where the sites $\beta$ are the nearest neighbors of $\alpha$.  We
include a factor of $1/\eta^4$ compared to Eq.~\ref{Hising} for
convenience.  In terms of the link degrees of freedom, the extra term
is simply a loop tension $-h\sum_p \sigma^z_p$. The other diagonal
term in $H$ can also be thought of as a loop tension, but fine-tuned
to make the ground state have the weighting of the 2d classical Ising
model. 

The Hamiltonian (\ref{eq:h2}) reduces to that of the 2D
transverse field Ising model on the square lattice for $\eta=1$. There
is a phase transition at $h\approx0.32847$,\cite{blote} which is in
the 3D Ising universality class with the dynamical critical exponent
$z=1$ and the correlation length exponent $\nu\approx0.632$.  From the
discussion in the previous section, we know that there is a conformal
critical point at $h=0$ and $\eta=\eta_c$. This phase transition is in
the 2D Ising universality class with the dynamical critical exponent
$z\approx2.167$ and the correlation length exponent $\nu=1$.


\subsection{Phase diagram}

\begin{figure}[t]
\includegraphics[width=\columnwidth]{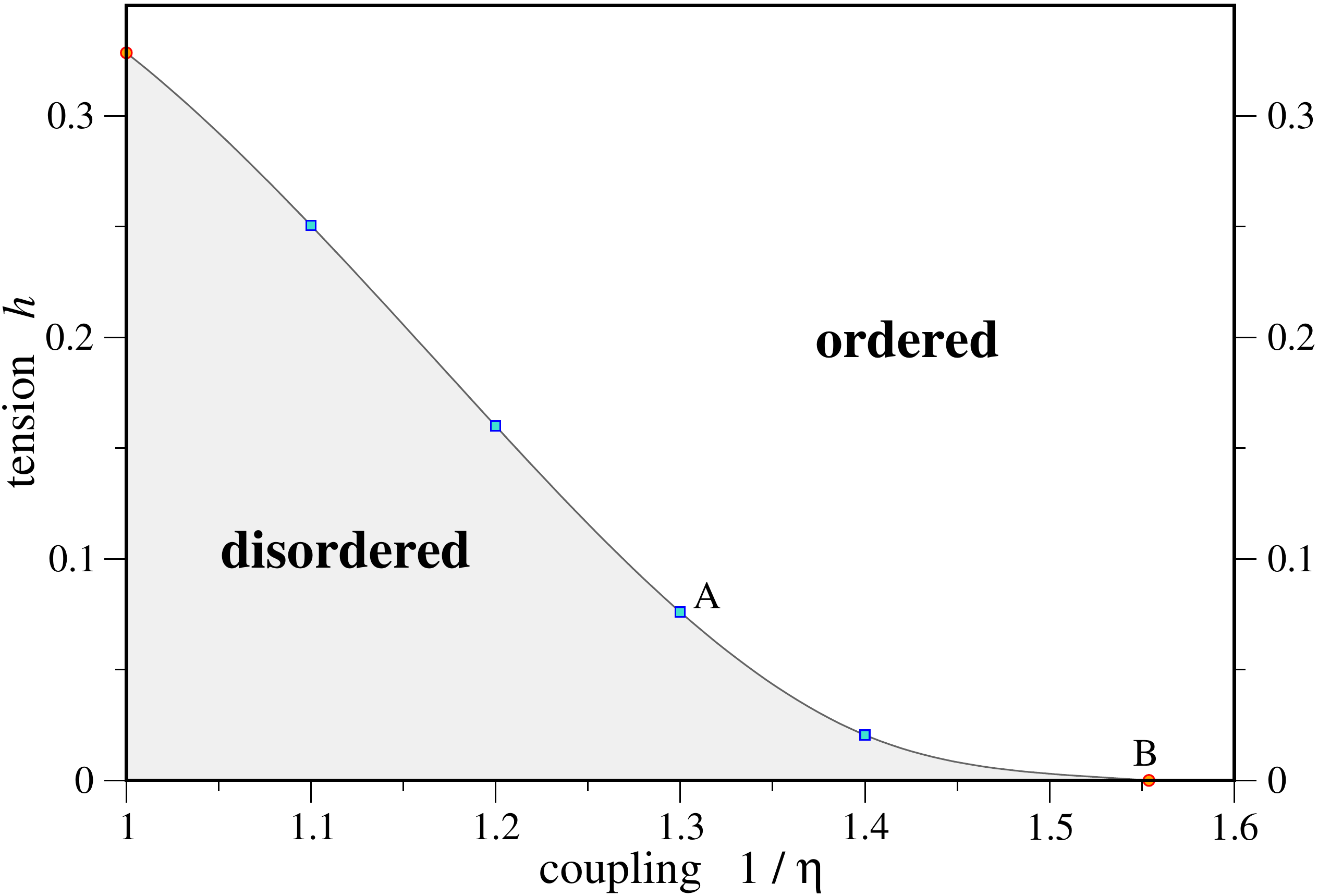}
\caption{Phase diagram in the $h$-$1/\eta$ plane. Error bars are smaller than
the symbol size. The line guides the eye. The whole transition line has $z=1$
except for one point at $h=0$ and $1/\eta=(1+\sqrt{2})^{1/2}$. 
For the two points indicated as A and B, we extract the critical exponent $\nu$
from data collapses of the Binder cumulants in Figs.~\ref{fig:collapse} and \ref{fig:collapse:2},
respectively.}
\label{fig:phase:diagram}
\end{figure}

The interesting question is if these two critical points are connected
by a line of phase transitions, or describe disconnected regions in
parameter space. We find convincing numerical evidence that the former
is true. The full phase diagram in the $h-1/\eta$ plane is shown in
Fig.~\ref{fig:phase:diagram}. The CQCP at $h=0$ and $\eta=\eta_c$ is
unstable under perturbations by both $h$ and $\eta$. Along a
particular line in this plane, there is a flow from the Ising CQCP to
the 2+1d Ising fixed point. The whole phase boundary in
Fig.~\ref{fig:phase:diagram} has 3D Ising exponents except for one
point.

This phase diagram was mapped out by using a variant of the
continuous-time quantum Monte Carlo algorithm.\cite{rieger} We measure
the Binder cumulant in Monte Carlo simulations
\begin{equation}
  U=1-\frac{\langle m^4\rangle}{3\langle m^2\rangle^2} \,,
  \label{eq:Binder}
\end{equation}
where $m$ is the magnetization density. The Binder cumulant scales in the
vicinity of a continuous phase transition as
$$
 U(L,K,\beta)=F(L^{1/\nu}(K-K_c),\beta/L^z),
$$
where $F$ is the scaling function, $L$ is the linear system size, $z$ is
the dynamical critical exponent, $\nu$ is the correlation length exponent,
$K-K_c$ is the distance to the critical point in some coupling constant $K$,
and $\beta$ is the inverse temperature. It follows from the above equation
that the curves for different system sizes should collapse onto the universal
curve $F$ for appropriate values of $\nu$ and $K_c$ when $\beta/L^z$ is fixed.
We locate the critical points shown in Fig.~\ref{fig:phase:diagram} by
collapsing the Binder cumulant data.

\begin{figure}[t]
\includegraphics[width=\columnwidth]{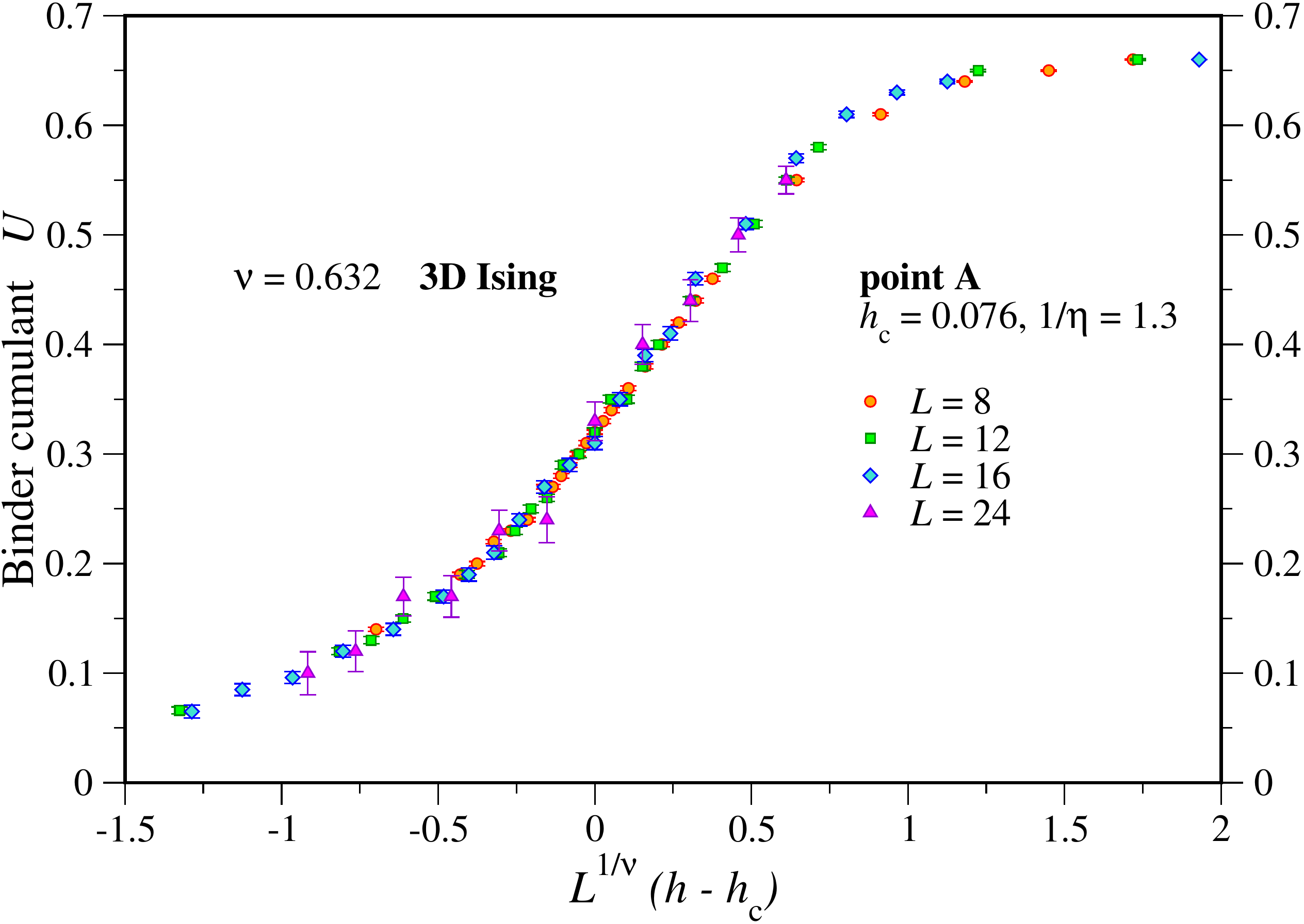}
\caption{Data collapse of the Binder cumulant \eqref{eq:Binder} for point A at $h_c=0.076$, $1/\eta=1.3$
using the 3D Ising exponent $\nu=0.632$ ($z=1$).
Data is obtained from continuous-time Monte Carlo simulations at inverse temperature $\beta=2L$.
}
\label{fig:collapse}
\end{figure}

\begin{figure*}[t]
\includegraphics[width=0.95\columnwidth]{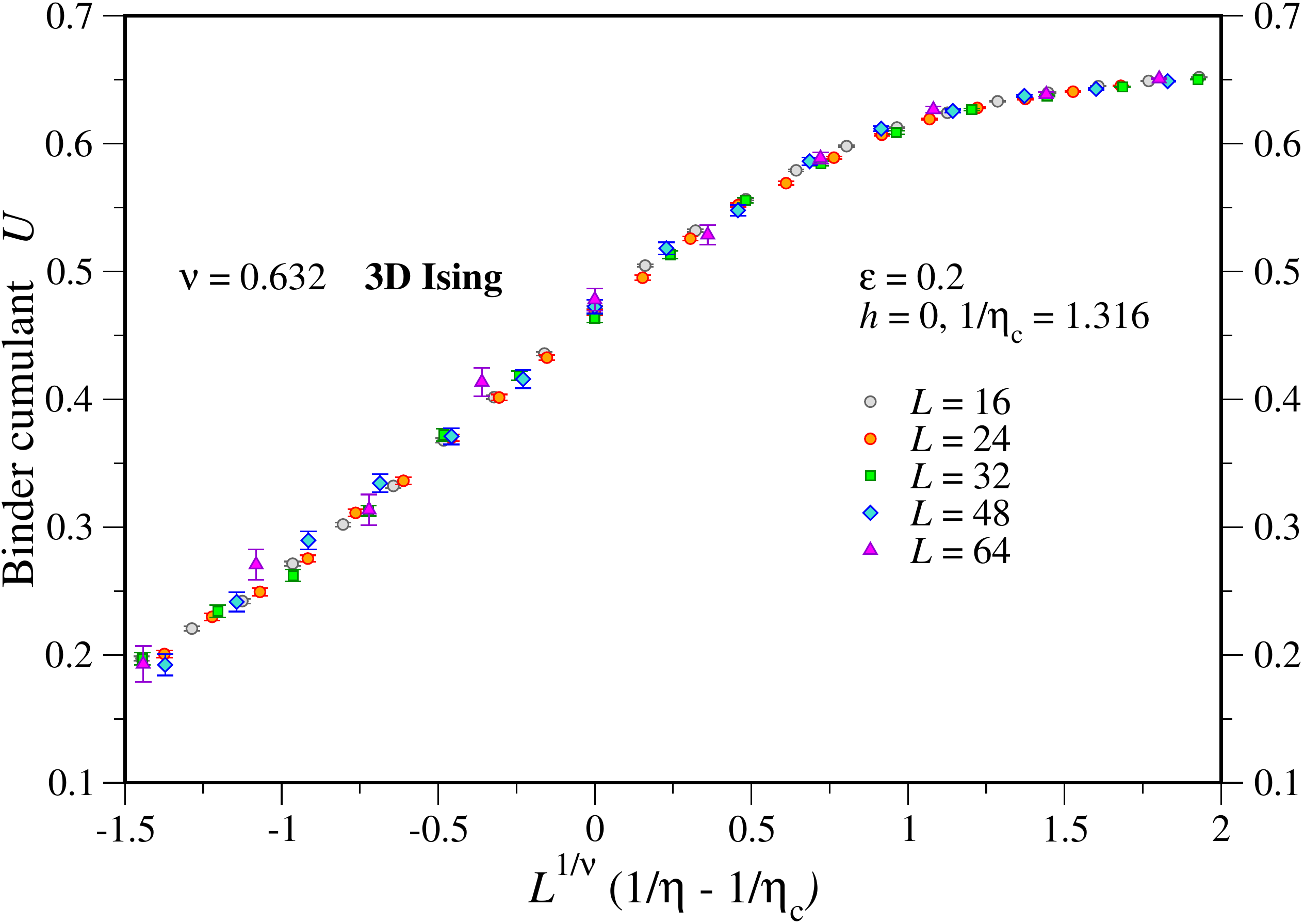}
\hskip 0.1\columnwidth
\includegraphics[width=0.95\columnwidth]{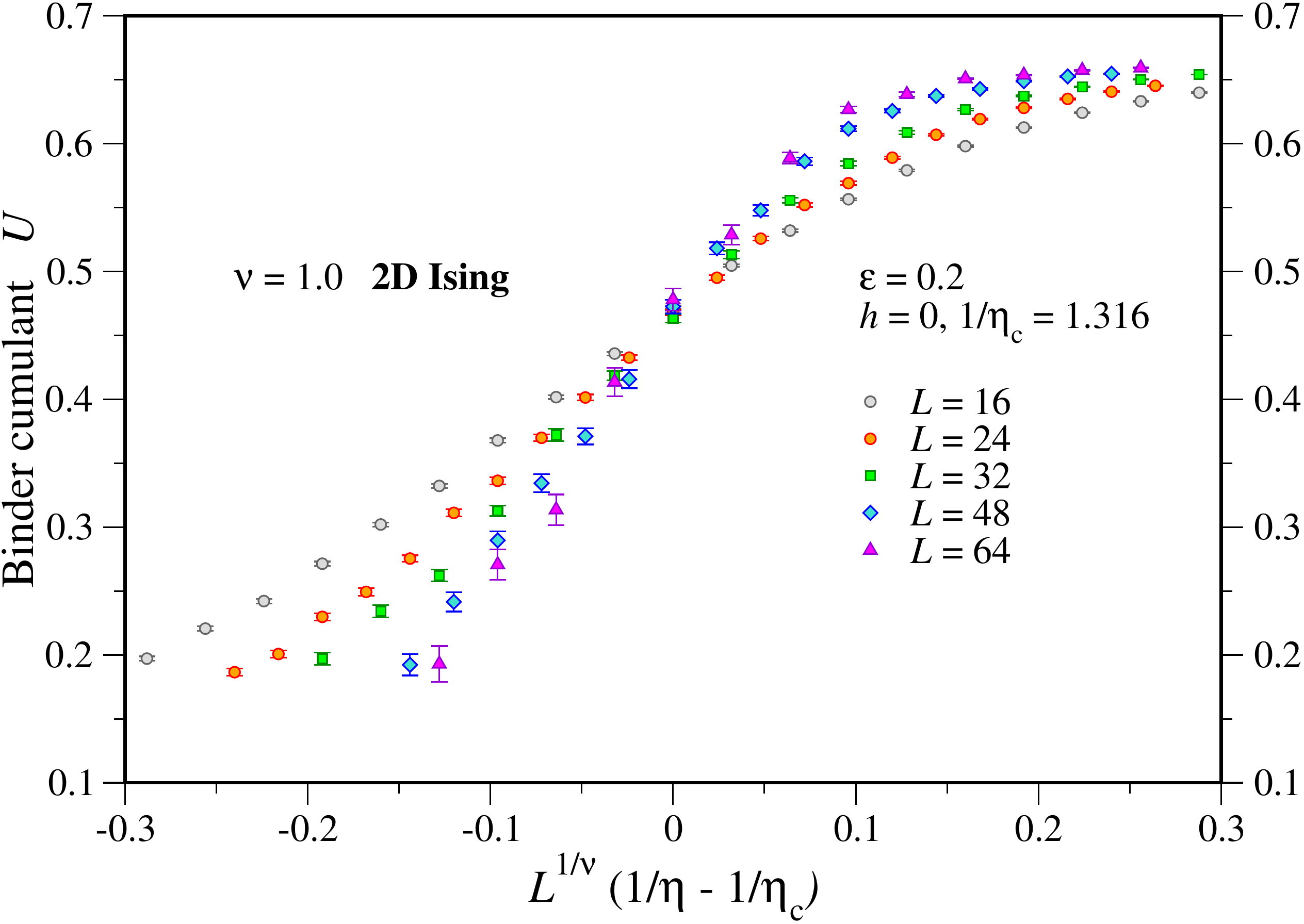}
\caption{Data collapse of the Binder cumulant \eqref{eq:Binder} for $h=0$ and $(1/\eta)_c=1.316$,
obtained from discrete-time Monte Carlo simulations with $\epsilon=0.2$ and
$\beta=3$.
Left panel: 3D exponent $\nu=0.632$. Right panel: 2D exponent $\nu=1$.
}\label{fig:collapse:discrete}
\end{figure*}

In Fig.~\ref{fig:collapse}, we show an example of such data collapse
for $h_c=0.076$, $1/\eta=1.3$. It is very hard to obtain the phase
boundary close to the Ising CQCP using the continuous-time algorithm
because the gap becomes very small and one needs to simulate very low
temperatures to obtain any meaningful results. We believe that the
phase transition is in the 3D Ising universality class with $z=1$ and
$\nu=0.632$ at any point on the phase boundary for finite values of
$h$.


\subsection{Instability of the 2D Ising point to Trotter errors}

In this subsection, we show that the Ising CQCP is unstable even to
the Trotter discretization errors. The imaginary time direction can be
discretized in $N=\beta/\epsilon$ slices and using the Suzuki-Trotter
decomposition, one maps the 2D quantum model onto the 3D (Ising-type)
classical model with inter-layer couplings given by the diagonal couplings
in Eq.~(\ref{eq:h2}) and the intra-layer ferromagnetic coupling
$K^\tau=-1/2\ln\tanh \epsilon\Gamma$, where $\Gamma=1/\eta^4$ is
the strength of the quantum transverse field. $K^\tau$ diverges in the limit
$\epsilon\to 0$.

This approach is often used to obtain the critical exponents when simulations
using the continuous-time algorithm are cumbersome. The exponents obtained
by discrete-time and continuous-time methods should have the same values
for stable fixed points. However, in our case, the 2D Ising point is unstable
and we obtain the 3D Ising exponents in discrete time indicating that there
is a flow to the 3D Ising fixed point, see Fig.~\ref{fig:collapse:discrete}. 
Thus we may have the case when quantum-to-classical mapping fails.
In principle, one should recover 2D exponents in the limit
$\epsilon\to 0$. We are unable to do this because the third
direction becomes very large for small $\epsilon$ and Monte Carlo simulations
become quite impractical.


\subsection{PIGS simulations}

\begin{figure}[b]
\includegraphics[width=\columnwidth]{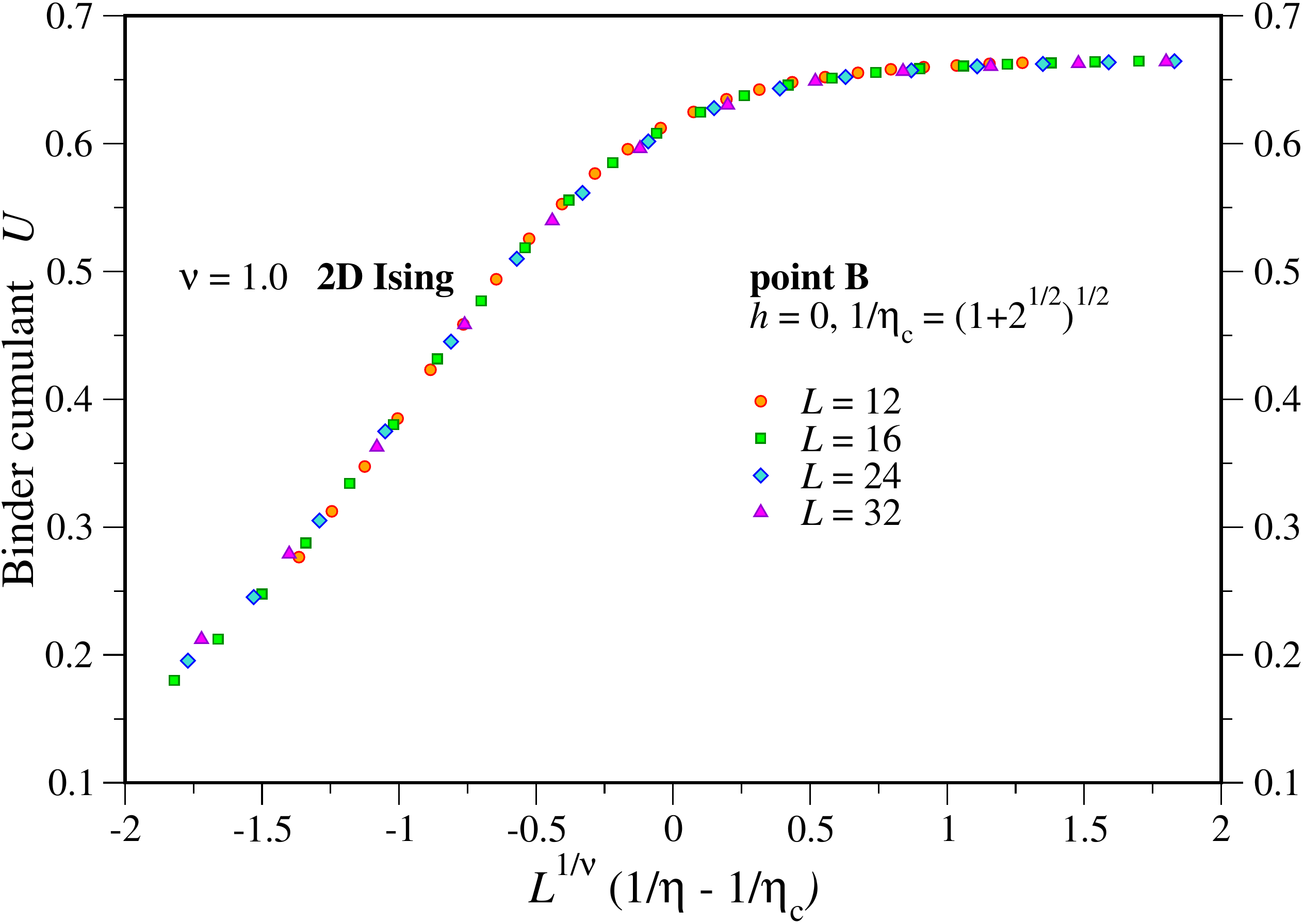}
\caption{Data collapse of the Binder cumulant \eqref{eq:Binder}  for point B at $h=0$,
$(1/\eta)_c=(1+\sqrt{2})^{1/2}$ using the 2D Ising exponent $\nu=1$.
The numerical data is obtained from continuous-time PIGS Monte Carlo simulations at $\beta=0.4$.}
\label{fig:collapse:2}
\end{figure}

The continuous-time quantum Monte Carlo simulations around the
Ising CQCP become very slow because we use local updates in the
spatial direction.  To prove that the transition at $h=0$ and
$\eta=\eta_c$ is indeed in the 2D Ising universality class, we perform
quantum Monte Carlo simulations by using the path integral ground
state (PIGS) Monte Carlo algorithm.\cite{pigs} In the PIGS algorithm,
one uses a variational wave function and tries to project the ground
state wave function. In our case, the ground state is known exactly
(strictly speaking, the wave function given in Eq.~(\ref{gs}) is not the
exact ground state in discrete time but becomes one in the limit
$\epsilon\to 0$). Thus simulations can be performed at any
temperature if one uses the continuous-time algorithm. We choose
$1/T=0.4$.  This temperature is low enough to have substantial quantum
dynamics and to avoid trivial classical simulations of the Ising model
that one effectively has at high temperatures.
Figure.~\ref{fig:collapse:2} shows the data collapse of the Binder
cumulant \eqref{eq:Binder} with the 2D Ising correlation length exponent. The 2D Ising
exponents can also be recovered using the PIGS Monte Carlo algorithm
in discrete time at small enough values of $\epsilon$, not shown.


\section{Conformal quantum critical lines with constant and varying $z$}
\label{sec:cqcp}

The two-dimensional classical Ising model at its critical point
provides a simple example of a conformal field theory. All the scaling
dimensions can be determined, and one finds that there are no marginal
perturbations possible. However, by coupling two critical Ising models
together, one finds a theory with an exactly marginal operator,
leading to a line of renormalization group (RG) fixed points.  If the
Ising spins are on the same lattice, this is called the Ashkin-Teller
model, while if they are on interpenetrating square lattices, this is
called the eight-vertex model.\cite{Baxbook} In equivalent fermionic
language, this amounts to coupling two Majorana fermions together with
an exactly marginal four-fermion term.

This richer behavior persists in quantum theories with RK
Hamiltonians. We review below how this coupling yields two conformal
quantum critical lines.\cite{AFF}
For any given point on one line one can map its correlators 
-- given in terms of the underlying classical theory --
to those of a point on the other line.
Thus the equal-time correlators in the quantum theory have
the same behavior at these two related points. 
This, however, is no guarantee that the quantum theories have the same 
dynamical behavior. Indeed we will show in the following that the dynamical 
behavior turns out to be quite different.

On one of these lines, the quantum six-vertex model, the model has a
height description and so has a $U(1)$ symmetry. Thus one expects the
quantum Lifshitz theory to describe the scaling limit, and indeed we
see numerically that $z$ remains $2$ all along this line. The second
conformal quantum critical line includes the case where the two Ising
CQCPs are decoupled. Thus there is no $U(1)$ symmetry, and as we saw
above, $z$ is not 2. 
However, these two critical lines intersect, and
so necessarily $z$ cannot remain at the Ising value along this line.
We present numerical work that indicates that $z$ indeed varies continuously
along this line.

In fact, an argument adapted from Ref.\ \onlinecite{AFF} suggests that
$z\ge 2$ for a CQCP. By relating the stress tensor of the
two-dimensional conformal field theory describing the ground state of
the CQCP to deformations of the quantum Hamiltonian, it is argued that
terms in the latter can only depend on space via the Laplacian, e.g.\
terms like $(\nabla^2\varphi)^2$ in the quantum Lifshitz theory. In
general there is no scalar field, but the argument still suggests that
the usual Lorentz-invariant terms in the effective action yielding
$z=1$ are absent. Moreover, fields in a unitary conformal field theory
necessarily have non-negative dimension, so acting with the
dimension-2 Laplacian gives a field of dimension at least 2. This
suggests that $z\ge 2$ for CQCPs where the underlying conformal field
theory is unitary, as for the theories considered here. It would be
interesting to make this argument more precise; we indeed find that
this bound is satisfied for all the CQCPs we study.

\subsection{The model}

\begin{figure}[t]
\begin{center}
\includegraphics[width=\columnwidth]{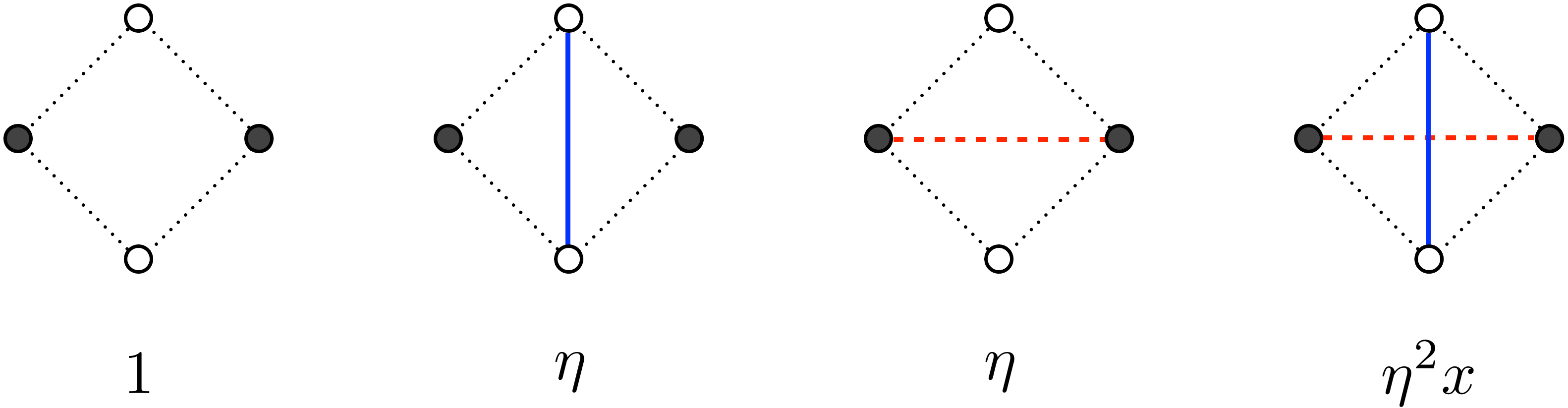}
\caption{The four configurations in the coupled deformed toric codes,
  and their weights in the ground state.}
\label{fig:toricsquared}
\end{center}
\end{figure}

As before, we study the Ising models in their domain wall/loop
formulation, so that the degrees of freedom live on the links of the
lattice. We place these models on two interpenetrating square lattices
(i.e.\ on both a square lattice and its dual).  Then the quantum model
has a four-state system on each face of the doubled lattice,
illustrated in Figure \ref{fig:toricsquared}.  The sites on the
original lattice are labeled by solid circles, while the dual lattice
sites are labeled by the open circles, so the figures applies to half
the faces; the others are given by a rotation. The constraint 
(\ref{constraint}) is required on all sites on the doubled lattice,
assuring that neither type of loop  branches or ends.

The phase diagram of the coupled model is well understood, because the
corresponding classical model is equivalent to the eight-vertex model
.\cite{Baxbook} This can be seen by first reverting to the original
Ising spins on the doubled lattice, and then drawing the domain walls
on the dual lattice to the doubled lattice. The four configurations in
Fig.~\ref{fig:toricsquared} then become the eight vertices of the
zero-field eight-vertex model. The counting is as follows. There are
of course 16 possible configurations of four Ising spins around a
plaquette. One can flip all the spins on the lattice of solid circles
without changing the configurations in Fig.~\ref{fig:toricsquared}.
One can do the same
for the lattice of open circles. Flipping the spins on both lattices does
not change the eight-vertex configuration, but flipping the spins on
say the solid lattice does. Thus the 16 different Ising spin
configurations correspond to the eight different configurations in the
eight-vertex model and the four different configurations in
Fig.~\ref{fig:toricsquared}. In the
standard eight-vertex model language, the four configurations in
Figure \ref{fig:toricsquared} have weights $c,a,b$ and $d$, respectively. 
To make the connection with coupled Ising domain walls more apparent,
we set $c=1/\eta$ and $d=\eta x$ and $a=b=1$.

The RK Hamiltonian with this ground state is therefore the same as that
of Ref.\ \onlinecite{AFF}. It is comprised of a sum over 2 by 2 blocks
just like the Ising case. The
off-diagonal terms flip all the solid lines around a plaquette on the
dual lattice, or flip the dotted lines around a plaquette on the
original lattice. The 2 by 2 block thus acts on four faces surrounding
each point on the doubled lattice. Let $n_c$ be the number of these
four faces which are empty, and $n_d$ be the number of crossings in
these four.  Likewise, let $\widetilde{n}_c$ and $\widetilde{n}_d$ be
the number of empty faces and crossings in the {flipped}
configuration.  Then the diagonal term for each site is given
by $c^{2\widetilde{n}_c} d^{2\widetilde{n}_d}$, i.e.\ what enters is
the number of crossings and empty faces in the {\em flipped}
configuration. Each 2 by 2 block is then 
\begin{eqnarray}
\mathcal{Q}_i &=& c^{\widetilde{n}_c + n_c} d^{\widetilde{n}_d + n_d}
\begin{pmatrix} 
c^{\widetilde{n}_c - n_c} d^{\widetilde{n}_d - n_d} & -1 \\
-1 & c^{n_c-\widetilde{n}_c } d^{n_d - \widetilde{n}_d} \\
\end{pmatrix}  \nonumber\\
&=&
\begin{pmatrix} 
c^{2\widetilde{n}_c} d^{2\widetilde{n}_d} & - 
c^{n_c+\widetilde{n}_c} d^{n_d+\widetilde{n}_d} \\
-c^{n_c+\widetilde{n}_c} d^{n_d+\widetilde{n}_d}
 & c^{2n_c} d^{2n_d} \\
\end{pmatrix}
\,,
\label{eq:qi}
\end{eqnarray}
where $c=1/\eta$ and $d=\eta x$ as before. The reason for the
prefactor as compared to Ref.\ \onlinecite{AFF} is to ensure that the
terms remain finite in the $c\to 0$ and $d\to 0$ limits. The fact that
$n_c+n_d + \widetilde{n}_c +\widetilde{n}_d =4$ can be used to rewrite
the off-diagonal terms if desired.

\subsection{Constant $z=2$ along the $U(1)$ symmetric line}

One critical line occurs when there is a $U(1)$ symmetry, on the
quantum six-vertex line. This $U(1)$ symmetry arises when $x=0$, i.e.\
the dotted and solid lines are never permitted to cross in the ground
state, so that there is a three-state system on each face.  The
symmetry becomes apparent by rewriting the degrees of freedom in terms
of an integer-valued ``height'' $h$ at each site of doubled lattice.
Heights on adjacent sites differ by one. The solid lines in Fig.
\ref{fig:toricsquared} represent domain walls between heights on the
original lattice, while the dashed lines are domain walls between
heights on the dual lattice. On a disc or sphere, this definition is
unique up to an overall shift $h\to h+n$ or a sign flip $h\to
-h$. (When $x\ne 0$, one can consistently assign only ${\mathbb
Z}_2$-valued heights, i.e.\ Ising variables, consistently.)

Thus this height has the same properties as the field
$\varphi$ in the quantum Lifshitz model discussed in Sec. \ref{Sec:Lifshitz}, 
and it is natural to identify $h$ with $\varphi$ in the continuum limit. 
However, as with the XY model, the classical model is not critical for all values of
$\eta$; a Kosterlitz-Thouless transition occurs at $c=2$. Thus for 
$x=0$, the model is critical for any
$\eta \ge 1/\sqrt{2}$. This critical line should
be described by the quantum Lifshitz model. The exponents in
equal-time correlators will depend on $\eta$, but we should have
$z=2$ all along this line. We have checked this explicitly for one point
at  $x=0$ and $\eta=1/\sqrt{2}$.

A subtlety in taking the $x=0$ limit is that crossings are not
forbidden with this Hamiltonian. Rather, they become fixed
defects. However, except in peculiar special cases all such defects
have a non-zero gap, and so can be ignored in the scaling
limit. Another thing to note is that there are extra ground states on
the torus in the $x=0$ limit. These are the analog of tilted states in
the dimer case, where the height is not periodic around a cycle of the
torus. In the language of dashed and solid loops, these result from
configurations where the loops around a cycle alternate between dashed
and solid. Two neighboring loops of the same type around a cycle can
annihilate, but loops of different type cannot: they lie on different
sets of links. Thus when crossing is forbidden, there is no way for
non-contractible loops of alternating type to annihilate.

\subsection{Continuous variation of $z$ along the ${\mathbb Z}_2$-symmetric line}
\label{sec:varyz}

\begin{figure}[b]
\includegraphics[width=0.8\columnwidth]{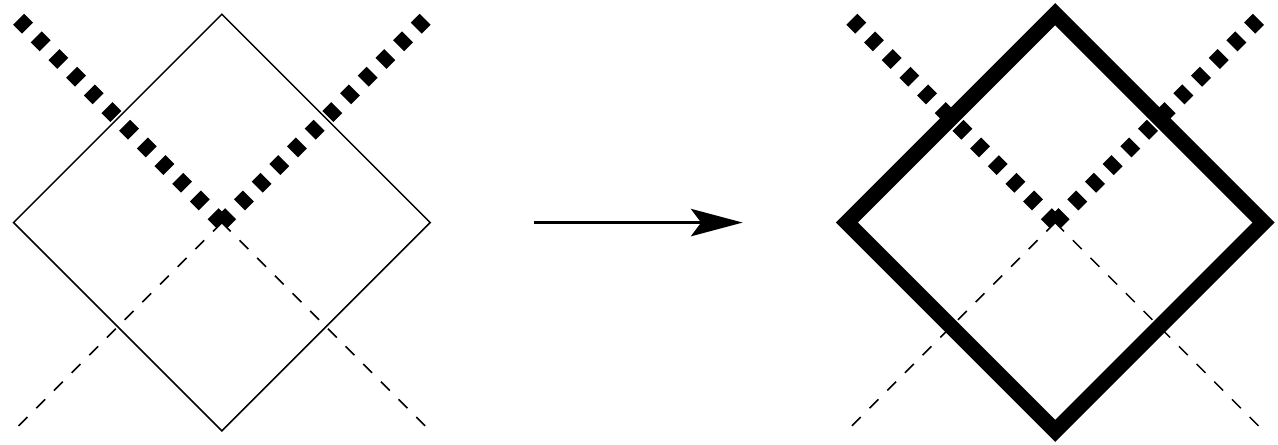}
\caption{Two configurations differ by a single plaquette flip. Thin lines
denote empty faces and thick lines denote nonempty faces. Here $n_c=2$,
$n_d=0$, $\tilde{n}_c=0$, and $\tilde{n}_d=2$, see text.}
\label{fig:plaq}
\end{figure}

The second critical line\cite{AFF} in the quantum eight-vertex model,
which does not have a $U(1)$ symmetry, is parametrized by
$c^2=d^2\pm 2$, or
\begin{equation}
x_c^2 = \frac{1}{\eta^4} \mp \frac{2}{\eta^2}\ .
\label{critline}
\end{equation}
We will refer to this line as the ``${\mathbb Z}_2$ critical line".
For $x_c=0$ this line meets the $U(1)$ critical line discussed in the 
previous section, and we thus expect a dynamical critical exponent
$z=2$ at this point.
However, when $\eta^2=\sqrt{2}-1$ on the ${\mathbb Z}_2$ line, we recover the case of two
decoupled Ising models corresponding to $x_c=1$. 
For this second point we thus expect a dynamical critical exponent of $z \approx 2.17$ 
as previously obtained in extensive numerical simulations of the dynamics of the 2d Ising
model,\cite{LandauBinder} see also Fig.~\ref{fig:ising:gap}.
As a result, one is led to expect that the dynamical critical exponent is varying 
(continuously) along the ${\mathbb Z}_2$ critical line.
This might not be surprising since it is known that all static critical exponents 
(of the equal-time correlators) are continuously varying along the ${\mathbb Z}_2$ and $U(1)$ critical lines. 
Thus it might be natural to expect that in the absence of additional symmetries
all critical exponents, including $z$, are continuously varying along such a critical line.

\begin{figure}[b]
\includegraphics[width=\columnwidth]{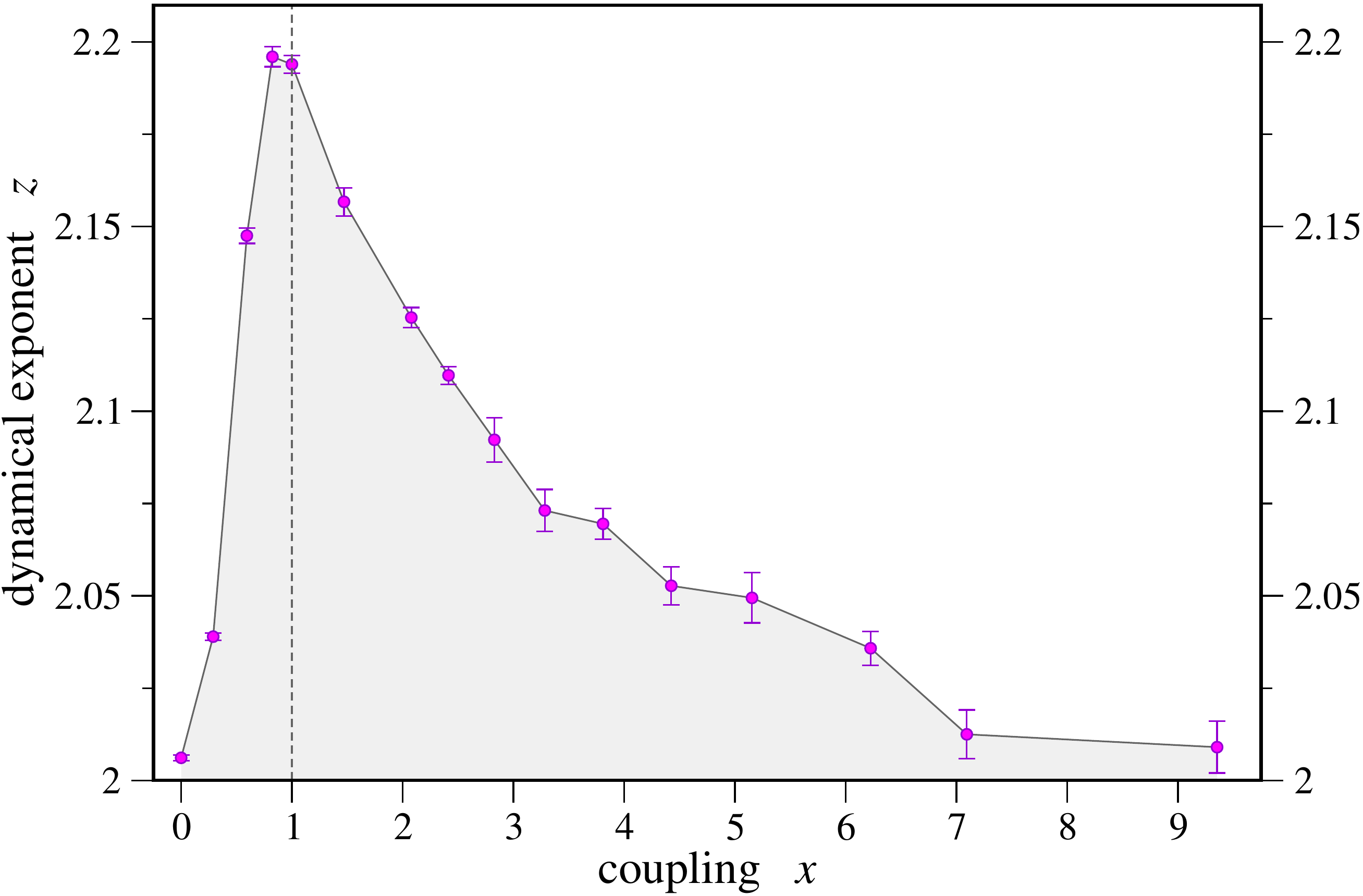}
\caption{The dynamical exponent $z$ as a function of $x$ along the critical
line. The dashed line indicates the coupling $x=1$ of the case of two decoupled 
classical Ising models. Error bars shown correspond to the statistical errors of 
the numerical fit described in Appendix \ref{Sec:Numerics}. 
For a discussion of systematic errors see Appendix \ref{Sec:DecouplingPoint}.
}
\label{fig:zx}
\end{figure}

To calculate the variation of the dynamical critical exponent along the ${\mathbb Z}_2$ critical line, 
we again perform classical Monte Carlo simulations.
The off-diagonal part of the Hamiltonian of the two coupled deformed toric
codes is more complicated compared with the single deformed toric code.
To work out the transition matrix, we note that the weights before and
after update are
$
  w({\cal C})=(x\eta^2)^{2n_d}\eta^{2\tilde{n}_c+2\tilde{n}_d}
$
and 
$
  w({\cal C}')=(x\eta^2)^{2\tilde{n}_d}\eta^{2n_c+ 2n_d} \,,
$
respectively. An example of such an update is shown in Fig.~\ref{fig:plaq}.
Given the off-diagonal matrix element
$
  Q_{ij}=-(1/\eta)^{n_c+\tilde{n}_c} (x\eta)^{n_d+\tilde{n}_d},
$
it is easy to obtain the transition matrix
$$
  W_{{\cal C},{\cal C}'}=w({\cal C}').
$$
This is indeed a legitimate transition matrix since $0\le w\le 1$.

In Fig.~\ref{fig:zx}, we show our numerical estimates of the dynamical
critical exponent $z$ as a function of the coupling $x$. The point
$x=0$ is the special point where the two critical lines meet. Here
$z=2$ (within error bars), as it remains all along the critical line
with $U(1)$ symmetry ($x=0$ and $\eta \ge 1\sqrt{2}$). Moving along
the ${\mathbb Z}_2$ critical line with increasing coupling $x$, we see
that $z$ indeed varies continuously. The dynamical exponent has a
non-monotonic behavior with an increase up to a maximum of $2.196$
close to the Ising point ($x=1$) and a decrease down towards $z=2$ as $x$
goes to infinity. We note that there might be small systematic errors,
which are discussed in Appendix \ref{Sec:DecouplingPoint} in detail,
because the system sizes are not large enough for the majority of points
in Fig.~\ref{fig:zx}.

The contrasting observation that the dynamical critical exponent does
not change along the $U(1)$ critical line -- despite continuously
varying equal-time correlators -- suggests that this additional $U(1)$
symmetry protects the dynamical critical exponent $z=2$.  This is in
line with the general classification of Hohenberg and Halperin
,\cite{HH} within which the ${\mathbb Z}_2$ symmetric line is in the
{\em Model A} dynamical symmetry class (absence of any conservation
law), and the $U(1)$ symmetric line in the {\em Model B} dynamical
symmetry class (presence of a conservation law).


\section{Discussion}

We have studied the dynamics of a number of conformal quantum critical
points. We explored the Ising CQCP in depth, showing that its dynamics
are equivalent to classical dynamics. We also showed that it is quite
unstable, and that by perturbing with the usual nearest-neighbor
interaction it can be continuously connected to the conventional
Lorentz-invariant Ising transition in 2+1 dimensions.  By studying two
coupled deformed toric codes, we illustrated how different dynamics
can result in different values of $z$, even when the equal-time
correlators are the same.

There are several interesting directions for future research.  At the
beginning of Sec. \ref{sec:cqcp} we discussed two quantum-critical
lines, one possessing $U(1)$ symmetry and the other possessing only
${\mathbb Z}_2$ symmetry.  These two lines intersect at one point at
which the dynamical critical exponent is $z=2$.  This point, not
surprisingly, has enhanced symmetry. For example, the equal-time
correlations are known to possess an $SU(2)$ symmetry. Moreover, the
physics should be in the same universality class as the loop models
studied in Ref.\ \onlinecite{Troyer08}. The coupling constant moving
one along the ${\mathbb Z}_2$ line away from this intersection point with $SU(2)$
symmetry must be an exactly marginal perturbation which generates a
line of fixed points for both statics and dynamics. In particular, by
performing perturbation theory in this coupling constant around the
$SU(2)$ point, one expects to be able to obtain analytic expressions
for the deviation of the dynamic critical exponent from $z=2$. 

It would be interesting to explore whether the quantum critical line recently 
postulated \cite{Vidal10} for the toric code model in a multicomponent 
magnetic field bears some relation to our results. It was found \cite{Vidal10}
that along this line the product $z \nu$ of dynamical critical exponent 
and correlation exponent appears to vary from approximately $0.69$ to $1$.
The most likely interpretation of those results might be in terms of a crossover 
between a conformal quantum critical point with $z=2$ 
and correlation exponent $\nu=1/2$ -- corresponding to the $x \to \infty$
limit of the ${\mathbb Z}_2$ critical line in the quantum eight-vertex
model studied in this manuscript -- and a Lorentz-invariant (2+1)
dimensional multicritical point. Such a scenario would be akin to 
the crossover discussed in detail in Sec. \ref{Flow2Dto3DIsing}.

One may further ask if it is possible to define a quantum dynamics of RK type
based on classical stochastic dynamics as discussed in this paper, for
(2+1) dimensional systems supporting {\it non-Abelian} statistics.  A
case in point is the Levin-Wen model,\cite{LevinWen} which can be
viewed\cite{NaturePhysicsPaper,BurnellSimon} as a `lattice
regularization' of (2+1)-dimensional doubled SU(2) Chern-Simons theory
at level $k=3$, possessing anyon excitations with non-Abelian exchange
statistics.  Indeed, the ground state $|\Psi\rangle$ of this model in
the simplest non-abelian (Fibonacci) case can be described in a
geometric form similar to Eq.~\eqref{gs}. Such a description is found by
using the results of Ref.\ \onlinecite{Fidkowski} to rewrite the Fibonacci
Levin-Wen model in terms of the quantum net model of
Refs.\ \onlinecite{FendleyFradkin,FendleyPaperNets}.  Here, the
configurations ${\cal C}$ describe configurations of so-called "nets",
which provide an orthonormal basis as in Eq.~\eqref{OrthoNormal}. 
A key difference with Eq.~\eqref{gs} is that $\sqrt{w({\cal C})}$ is replaced
by a real wavefunction of the configurations ${\cal C}$ which takes on
both positive {\it and} negative values.  (The lack of positivity of
the wavefunction ("sign problem") may be a general feature of systems
supporting excitations with non-Abelian statistics.)  This property
prevents one from constructing a quantum dynamics of RK type for the
Fibonacci Levin-Wen model by using a stochastic relaxational dynamics
of a suitable classical statistical model, as we did in Sec.
\ref{sec:stochastic}.

Nevertheless, there is a classical 2D statistical mechanics model
which arises naturally from the ground state $|\Psi\rangle$ of the
Fibonacci Levin-Wen model, and gives rise to a CQCP.\cite{FendleyFradkin,FendleyPaperNets}  
By tuning the weight per unit
length of net just as we did in this manuscript, the (2+0)-dimensional
classical partition function obtained from the wave function
$|\Psi\rangle$, $Z = \langle \Psi |\Psi\rangle$ is critical.  $Z$ is
now a sum of non-negative Boltzmann weights because only the square of
the wavefunction appears due to the orthonormality of the basis. By
Hastings' theorem, the quantum model must therefore be critical.  This
classical critical point is in the universality class of the
$(2+0)$-dimensional conformal minimal model with central charge
$c=14/15$ (or in the language of
Ref.~\onlinecite{FendleyFradkin,FendleyPaperNets}, the $Q$-state Potts
model with $Q=(5+\sqrt{5})/2$).\cite{FootnoteNaturePhysOnePlusOne}
Such a conformal field theory is
known\cite{ZamolodchikovMulticritical} to be described by a classical
Boltzmann weight arising from Landau-Ginzburg theory, precisely of the
type described above Eq.~\eqref{HamiltonianFokkerPlanck}.  Were it not
for the "sign problem" mentioned above, one could proceed to study the
dynamic critical exponent $z$ for this conformal quantum critical
point along the lines of the present manuscript.


\begin{acknowledgments}

We acknowledge discussions with E.~Ardonne.  This work was supported
by the Swiss National Science Foundation, the Swiss HP$^2$C initiative, NSF grants DMR/MSPA-0704666 (P.F.)  and NSF DMR-0706140 (A.W.W.L).
We thank the Aspen Center for Physics for hospitality both at the
initial and the final stages of this work.
S.T. acknowledges hospitality of the Kavli Institute for Theoretical Physics supported
by NSF PHY-0551164.

\end{acknowledgments}


\appendix

\section{Numerical calculation of the dynamical critical exponent}
\label{Sec:Numerics}

\begin{figure}[t]
\includegraphics[width=\columnwidth]{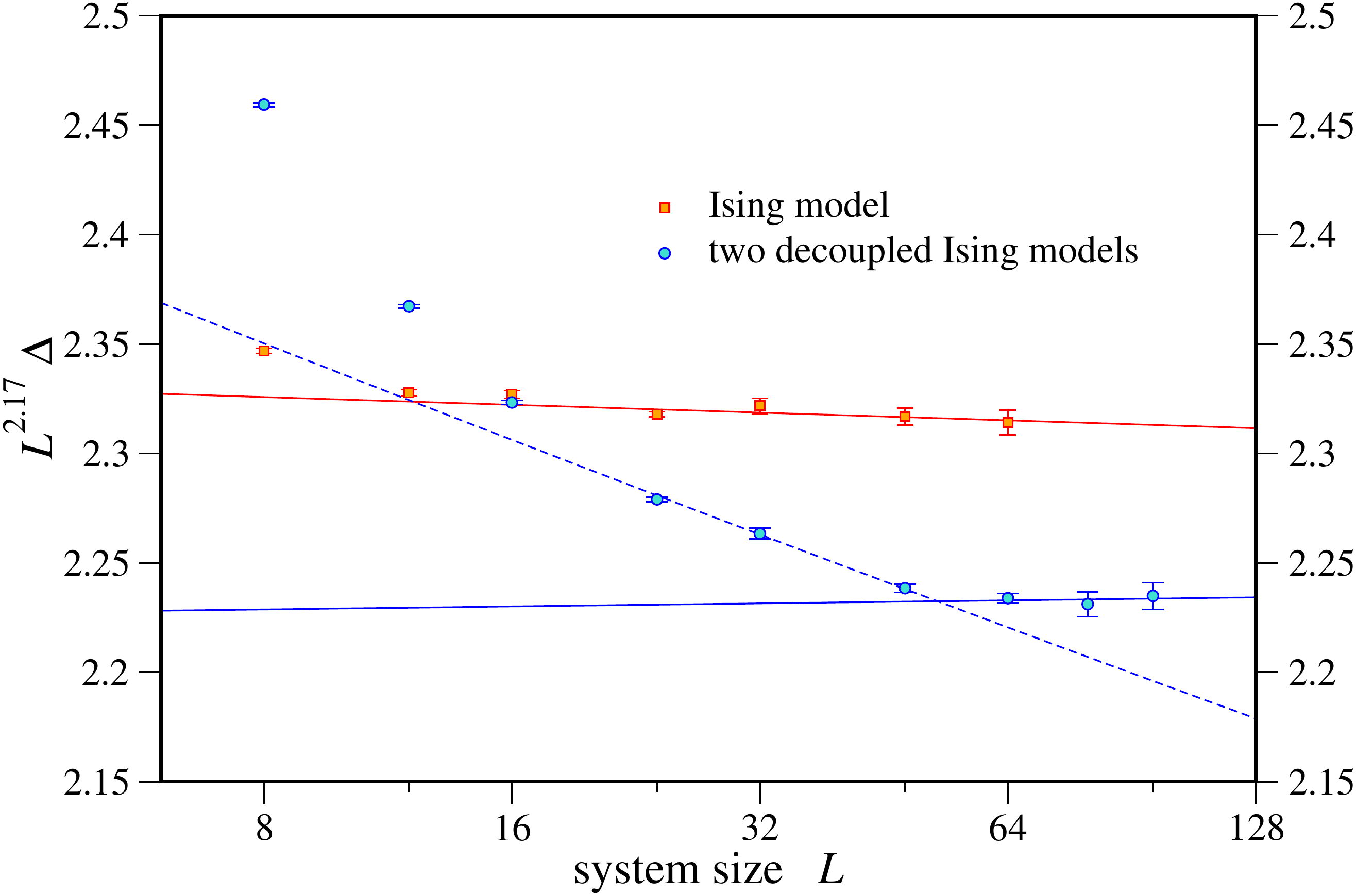}
\caption{Four times the gap of Hamiltonian~(\ref{Hising}) and the gap of
Hamiltonian~(\ref{eq:qi}) ($x=1$) at the critical point $\eta_c$ as
a function of the system size. The gaps are multiplied by $L^{2.17}$ to
make the difference between the two curves clearly visible. The gaps of
the two decoupled Ising models show very strong system size dependence.
Fitting the curve from $L=24$ to $L=64$ gives $z=2.192(3)$ (indicated by the
dashed line), 
whereas fitting
the curve from $L=64$ to $L=96$ gives $z=2.170(4)$ (indicated by the solid 
line)
as for the single Ising
model.}
\label{fig:one:two:gap}
\end{figure}

In this Appendix, we describe our numerical method. To extract
the dynamical critical exponent $z$, we use the finite size scaling of
the gap $\Delta\propto L^{-z}$
at the critical point, where $L$ is the linear system size. The gap can
be obtained from the autocorrelation function of classical Monte Carlo
simulations by ensuring that the off-diagonal part of the Monte Carlo
transition matrix is proportional to the off-diagonal part of a Hamiltonian
\cite{Henley97} (see also Sec. \ref{sec:stochastic}).
Strictly speaking, the Hamiltonian is symmetric but the
transition matrix $W_{{\cal C},{\cal C}'}$ is not. The following matrix is
symmetric and has the same eigenvalues as $W_{{\cal C},{\cal C}'}$:
\cite {Henley97}
$$
  \tilde{W}_{{\cal C},{\cal C}'}=
    w({\cal C})^{1/2} W_{{\cal C},{\cal C}'} w({\cal C}')^{-1/2},
$$
where $w({\cal C})$ and $w({\cal C}')$ are the statistical weights of
the configurations before and after update. If the transition matrix
$\tilde{W}$ is proportional to the Hamiltonian:
$$
  \tilde{W}=1-cH,
$$
where $c$ is the coefficient of proportionality, then the gap of the
Hamiltonian is related to the autocorrelation time $\tau_A$ of some
observable $A$ as
$$
  \Delta=\frac{1-e^{1/\tau_A}}{c},
$$
where $\tau_A$ is measured in Monte Carlo time units. The autocorrelation
time $\tau_A$ can be obtained by fitting the autocorrelation function to
an exponential function (at long enough Monte Carlo time to make sure that
the contribution from the other modes is negligible).

We perform Monte Carlo simulations as follows. We simulate a classical model
corresponding to the quantum Hamiltonian for different system sizes $L$.
Typically the largest system size is $L=64$. We measure the autocorrelation
time of the magnetization for Ising-like models and the autocorrelation time
of a product of two dimers separated by a distance of $L/2$ for the dimer
model. The dynamical exponent is calculated as described in the previous
paragraph. Typically we perform 25 independent runs for every point and
the error bars are obtained by using the jackknife method.


\section{Size dependence of the gap in the deformed toric code model}
\label{Sec:DecouplingPoint}

We note that at the decoupling point $x=1$ of the deformed toric code model,
describing two decoupled Ising models, the gap shows very strong system size
dependence reaching the asymptotic exponential form only at large
system sizes. In Fig.~\ref{fig:one:two:gap}, we show the gaps of this model
and the ordinary Ising model as functions of the system size. The dynamical
critical exponent of the two decoupled Ising models is $z=2.192(3)$ if we fit
the curve from $L=24$ to $L=64$ and $z=2.170(4)$ if we fit the curve from
$L=64$ to $L=96$. The latter value is in agreement with that for the ordinary
Ising model.\cite{Nightingale} 
The origin of this strong system size dependence might be tracked back to the
additional prefactor
$c^{\widetilde{n}_c + n_c} d^{\widetilde{n}_d + n_d}$ in Hamiltonian~(\ref{eq:qi}) 
as compared to the ordinary Ising Hamiltonian~(\ref{Hising}). 
This prefactor might give rise to slightly modified dynamics for small system
sizes.
Typically the largest system size in our Monte Carlo simulations is $L=64$
for small values of $x$ and $L=48$ for large values of $x$. Assuming that the
gaps for all points in Fig.~\ref{fig:zx} have strong system size
dependence, we may conclude that each point in Fig.~\ref{fig:zx} might have
a (small) systematic error.




\begin{thebibliography}{99}
\bibitem{FootnoteFrustration}
We note that Hamiltonians that are the sum of local projectors and
that have at least one zero-energy ground state are sometimes referred
to as ``frustration-free'', in particular in literature concerned with
quantum information. Distinguishing such Hamiltonians is
a good idea, since many interesting Hamiltonians can be written as a
sum over local projectors (e.g.\ the spin-1/2 Heisenberg model), but
typically there is no state annihilated by all of them. Unfortunately,
calling these ``frustration-free'' can lead to confusion, since part
of the physics interest in such models is that they often arise from
spin models with a large amount of frustration!  For example, the
classical dimer model on the honeycomb lattice is equivalent to the
zero-temperature antiferromagnetic Ising model on the triangular
lattice; the dimers themselves correspond to the frustrated bonds
.\cite{MS2} For this reason, we continue to refer to such Hamiltonians
as RK Hamiltonians.
\bibitem{FootnoteNaturePhysOnePlusOne}
Whereas this conformal field theory (CFT) describes
the  critical ground state wave function of the Fibonacci Levin Wen
model which resides in (2+0) dimensions, 
a conformal field theory
of the same central charge $c=14/15$ was found
by entirely different techniques
for the quantum dynamics of the same Levin Wen model
when spatially confined to quasi one dimensional geometry,
and tuned across a quantum critical point by  a
generic term in the Hamiltonian.  This CFT thus describes a
(1+1) dimensional system, and has dynamic critical exponent $z=1$,
analogous to the (2+1) dimensional Ising model described in 
Sec. \ref{Flow2Dto3DIsing}.
A priori, there is of course no obvious reason that the
same CFT should arise in these (2+0) and (1+1) dimensional
limits.
\end{thebibliography}
\end{document}